\documentclass[aps,twocolumn,prb,superscriptaddress]{revtex4}
\usepackage{amsfonts}
\usepackage[final,hiresbb]{graphicx}
\usepackage{amsmath}
\usepackage{amssymb}
\usepackage{amstext}
\usepackage{color}
\usepackage{braket}

\definecolor{Green}{RGB}{0,204,102}
\definecolor{Purple}{RGB}{102,0,255}
\definecolor{Blue}{RGB}{0,0,255}
\definecolor{Red}{RGB}{255,000,000}

\begin{document}

\title{Engineering and Manipulating Exciton Wave Packets}

\author{Xiaoning Zang}
\affiliation{Department of Physics, Colorado School of Mines, Golden, CO 80401, USA}
\author{Simone Montangero}
\affiliation{Institute for complex quantum systems \& Center for Integrated Quantum Science and Technology (IQST), 
Universit\"at Ulm, Albert-Einstein-Allee 11, D-89075 Ulm, Germany}
\affiliation{Theoretische Physik, Universit\"at des Saarlandes, D-66123 Saarbr\"ucken, Germany}
\author{Lincoln D. Carr}
\affiliation{Department of Physics, Colorado School of Mines, Golden, CO 80401, USA}
\author{Mark T. Lusk}
\email{mlusk@mines.edu}
\affiliation{Department of Physics, Colorado School of Mines, Golden, CO 80401, USA}

\begin{abstract}
When a semiconductor absorbs light, the resulting electron-hole superposition amounts to a uncontrolled quantum ripple that eventually degenerates into diffusion. If the conformation of these excitonic superpositions could be engineered, though, they would constitute a new means of transporting information and energy. We show that properly designed laser pulses can be used to create such excitonic wave packets. They can be formed with a prescribed speed, direction and spectral make-up that allows them to be selectively passed, rejected or even dissociated using superlattices. Their coherence also provides a handle for manipulation using active, external controls. Energy and information can be conveniently processed and subsequently removed at a distant site by reversing the original procedure to produce a stimulated emission. The ability to create, manage and remove structured excitons comprises the foundation for opto-excitonic circuits with application to a wide range of quantum information, energy and light-flow technologies. The paradigm is demonstrated using both Tight-Binding and Time-Domain Density Functional Theory simulations.
\end{abstract}


\maketitle

%
\begin{figure}[t]\begin{center}
\includegraphics[width=0.4\textwidth]{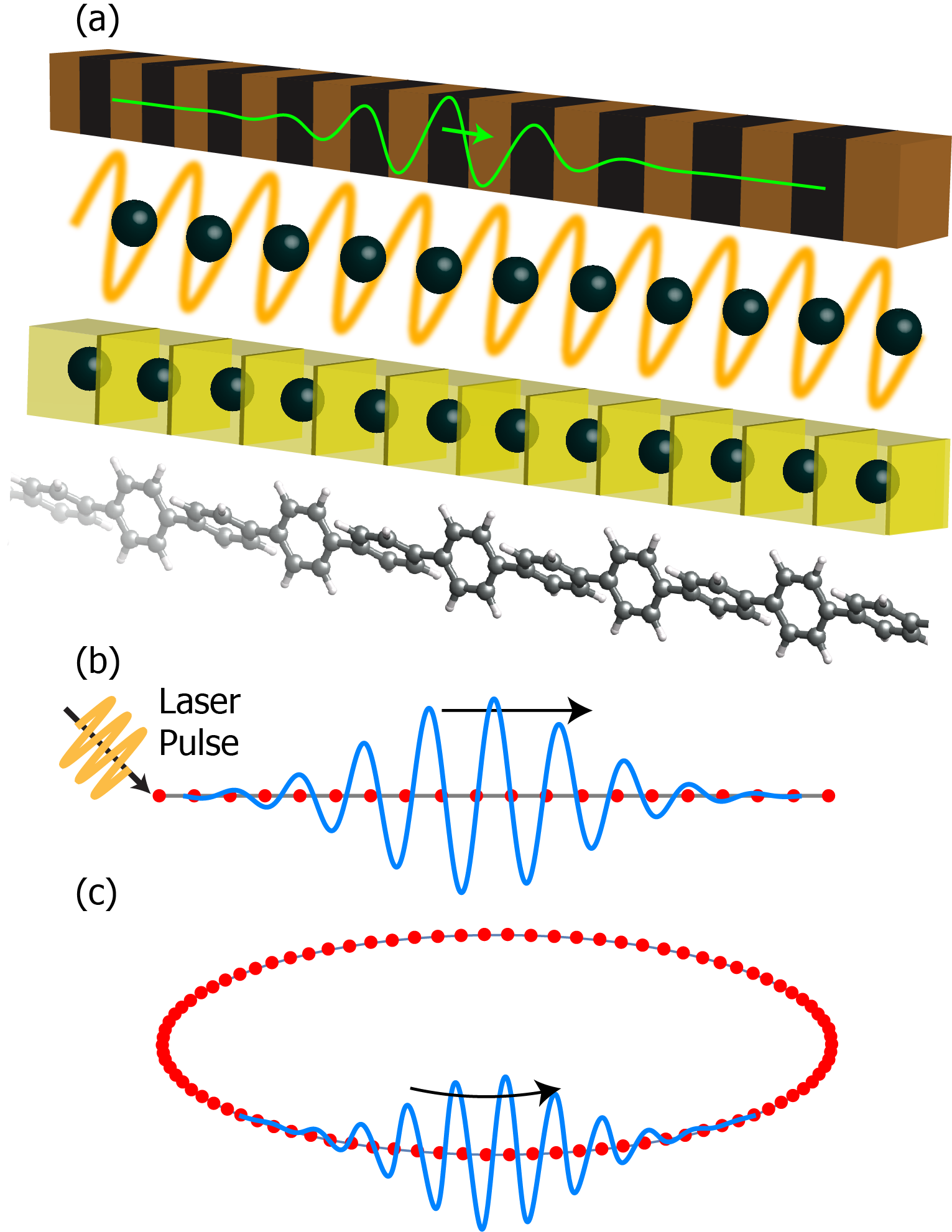}
\caption{
{\bf Material Settings for Quasi-1D Exciton Dynamics}. {\bf (a)} Possible implementations include inorganic quantum-well superlattices (top), Rydberg atoms in optical lattices (second), atoms in strongly coupled optical cavities (third), and organic molecular chains (bottom). {\bf (b)} Finite chain lattice of interest with photon absorption only at left-most site. {\bf (c)} Ring lattice construct used to design laser pulses. Sites are shown in red with structured excitons in blue.}
\label{geometries}
\end{center}
\end{figure}

\section{Introduction}
Although excitons are often thought of in association with diffusive energy flows~\cite{Mikhnenko_2015}, it is possible to characterize their dynamics prior to a loss of coherence~\cite{Hahn_PRB_1980}. For instance, unstructured superpositions have been identified using quantum beat spectroscopy~\cite{Stolz_PRL_1991, Feldmann_1993}, hot carrier luminescence spectroscopy~\cite{Elsaesser_PRL_1991}, emission spectra splitting~\cite{Bayer_Science_2001}, two-time anisotropy decay~\cite{Collini_Science_2009}, transport spectroscopy~\cite{Delbecq_NatComm_2013} and Hanbury-Brown and Twiss interferometry~\cite{Rivas}. Related excitonic Bloch oscillations have been measured using transient degenerate four-wave mixing~\cite{Leisching_PRB_1994}. Superpositions have even been generated at precise positions using high-energy electrons and quantified using cathodoluminescence~\cite{Yang_APL_2014}. Solar-generated superpositions have also received theoretical scrutiny as a possible means of increasing the efficiency of energy transport~\cite{Jang, Kreisbeck_2012} where partial entanglement with phonons actually makes them more robust in the face of disorder~\cite{Plenio_2008}. In all of these settings, though, the focus is on excitons that do not have a moving center or on the evolution of naturally occurring superpositions.

It is also possible to create ballistic excitons via photoexcitation. Inorganic II-VI quantum well superlattices support \emph{hot excitons}~\cite{Umlauff_1998} that relax through a well-characterized cascade of optical phonon emissions~\cite{Kalt_2005}. This allows excitons to be created with a known kinetic energy that spread isotropically within a layer until scattering with phonons. The spatial spread of their fluorescence over time can be precisely measured and shown to correspond to a constant speed of motion. This speed can even be controlled to some extent using the laser energy. Of course, this is ballistic spreading; the centroid of the exciton does not move in this quantum ripple phenomenon.

In contrast, it would be desirable to create spatially localized excitonic wave packets with a prescribed speed, direction and spectral content. The methodology could then be used to create opto-excitonic circuits as an alternative to photonic crystals~\cite{Yablonovitch_PRL_1987} wherein the flow of light is controlled while manifested as excitons instead of via the influence of local charge distributions. As with photonics~\cite{Krauss_Nature_1996}, fabrication methods could be borrowed from the semiconductor industry~\cite{Gartner}, but organic~\cite{Collini_Science_2009} and optical lattice implementations~\cite{Wuster_NJP_2011} may also be possible as illustrated in Fig. \ref{geometries}(a). Electronic and electrochemical technologies based on exciton dissociation carry this out using a material heterojunction with an inherent energy loss, but these excitonic wave packets can be dissociated using quantum interference without energy dissipation~\cite{Lusk_Fano_2015}. From the perspective of quantum information processing, they embody the mathematical formalism of Heisenberg spin packets and so would be able to store and transport quantum states~\cite{Osborne_PRA_2004, Haselgrove_PRA_2005}. However, the ease with which excitons can be manipulated adds ready qubit management to the spin chain paradigm~\cite{Bose_ContempPhys_2007, Thompson_2016, Seifnashri_2016}. All of these applications would benefit from an ability to create excitons with an engineered structure, and a methodology for doing so is the subject of this work. 

We initially treat excitons as indivisible particles to show how laser pulses can produce wave packets of prescribed shape and speed. This is subsequently generalized to consider exciton dissociation; the physics are richer when electrons and holes can move independently~\cite{Lusk_Fano_2015}. The basic idea of pulse shaping is most easily explained with a prescribed electric field and a two band Tight-Binding (TB) model. Real-time Time-Domain Density Functional Theory (RT-TDDFT) is then used to allow for multi-electron interactions and complex superpositions of  electron and hole states. Physically realizable laser pulses illuminating chains of organic molecules are shown to produce  structured excitons. This constitutes a proof-of-concept with fewer idealizing assumptions.  The effect of beam broadening and an account of exciton-phonon interactions within an open system setting are natural extensions~\cite{Ishizaki_2009}.

\section{Approach}
Attention is restricted to the quasi-one-dimensional settings of Fig. \ref{geometries}(a) which can be abstractly viewed as a series of lattice sites as shown in Fig. \ref{geometries}(b). The system Hamiltonian can be described in either a TB formalism or within the setting of RT-TDDFT. The former paradigm has the advantage of computational simplicity and a level of generality because it covers an entire class of material systems rather than a specific setting. RT-TDDFT, on the other hand, is used to provide time-explicit, quantitative information on the dynamics associated with a particular molecular assembly. After introducing both methodologies, we develop two supplementary techniques that prove useful in assessing the results of each method. One is an algorithm for post-processing RT-TDDFT data to estimate the time-evolving electron and hole populations on each site. The second analysis tool allows TB parameters to be determined directly from RT-TDDFT calculations. The suite of computational tools allows us to make quantitative comparisons between TB, RT-TDDFT, and analytical predictions for packet speeds. 

\subsection{Tight-Binding Model} 
Excitonic dynamics are considered within two types of Tight-Binding (TB) settings. The first is a single-particle model in which electrons are assumed to be either in a ground state or in a unique excited state on each lattice site.  Within a semi-classical approximation for applied electric fields, the single-particle TB Hamiltonian of interest is thus taken to be:
\begin{eqnarray}
\hat{H}  &=& \hat{H}_\Delta  +  \hat{H}_{\mathrm {ex}} +  \hat{H}_{\mathrm {laser}}, \nonumber\\
\hat{H}_\Delta  &=& \sum_{j}\Delta_{j}\hat{n}_{j}, \nonumber \\
\hat{H}_{\mathrm {ex}}   &=& \sum_{<i,j>,i\neq j}\chi_{i j}\hat{c}^{\dagger}_{j}\hat{c}_{i}  + \rm{H.c.},\label{H1} \\
 \hat{H}_{\mathrm {laser}} &=&  -\sum_{j} \bigl( \vec{\mu}_j \cdot \vec{E}\bigr) \hat{c}^{\dagger}_{j} + \rm{H.c.}\nonumber
\end{eqnarray}
$\hat{H}_\Delta$ is the band offset while $\hat{H}_{\mathrm {ex}}$ describes exciton hopping.  Roman subscripts $i$ and $j$ denote lattice sites, $<\!\!\!i,j\!\!\!>$ means a sum over sites that are nearest neighbors, $\hat{c}_{j}$ is the exciton annihilation operator for site $j$, $\hat{n}_{j} =\hat{c}^{\dagger}_{j}\hat{c}_{j}$ is the exciton number operator, and $[\hat{c}_{i}, \hat{c}^{\dagger}_{j}]=\delta_{ij}$. The energy of site $j$ is $\Delta_j$, the exciton hopping mobility is $\chi_{i j}$, and the lattice spacing is $a$. Phonon and photon coupling are disregarded. The transition dipoles at each site are given by $\vec{\mu}_j$, and the spatial variation of the electric field, $\vec{E}(t)$, is assumed to be negligible over the dimensions of interest---i.e. an electric dipole approximation is assumed.

The Hamiltonian of Eq. \ref{H1} can also be generalized to allow for distinct dynamics for electron and hole. This allows the consideration of both Frenkel and Wannier-Mott excitons and also makes it possible to study exciton dissociation. The requisite Hamiltonian of this two-particle TB model is:
\begin{eqnarray}
\hat{H}  &=& \hat{H}_\Delta  + \hat{H}_{e} +  \hat{H}_U  + \hat{H}_V +  \hat{H}_{\mathrm {laser}}, \nonumber \\
\hat{H}_\Delta  &=& \textstyle{\sum}_{n,\nu}\Delta_{n}^{\nu}\hat{n}^{\nu}_{n}, \nonumber \\
\hat{H}_{e}   &=& \textstyle{\sum}_{<m,n>,\nu}\chi^{\nu}_{m n}\hat{c}^{\dagger \nu}_{m}\hat{c}_{n}^{\nu} + \rm{H.c.},    \nonumber \\
\hat{H}_U   &=& \textstyle{\sum}_{n}U_{n}\hat{n}_{n}^1\hat{n}_{n}^2 + \rm{H.c.}, \label{H2}\\
\hat{H}_V   &=& \textstyle{\sum}_{m\neq n,\nu, \mu} V_{mn}^{\mu\nu}\hat{n}_{m}^{\mu}\hat{n}_{n}^{\nu} + \rm{H.c.}  \nonumber \\
 \hat{H}_{\mathrm {laser}} &=&  -\sum_{j} \bigl( \vec{\mu}_j \cdot \vec{E}\bigr) \hat{c}^{\dagger 2}_{j}\hat{c}_{j}^{1} + \rm{H.c.} \nonumber
\end{eqnarray}
Here $\hat{H}_e$ describes electron hopping while $\hat{H}_U$ and $\hat{H}_V$ are on-site and potentially long-range Coulomb interactions. Roman subscripts $m$ and $n$ denote lattice sites, Greek superscripts $\mu$ and $\nu$ indicate electron band, $<\!\!\!m,n\!\!\!>$ means a sum over sites that are nearest neighbors, $\hat{c}_{n}^{\nu}$ is the electron annihilation operator for band $\nu$ of site $n$, $\hat{n}^{\nu}_{n} =\hat{c}^{\nu\dagger}_{n}\hat{c}_{n}^{\nu}$ is the electron number operator, and $[\hat{c}^{\mu}_{m}, \hat{c}^{\nu\dagger}_{n}]_+=\delta_{mn}\delta_{\mu \nu}$.

Now consider the translational motion of an excitonic wave packet on the idealized ring geometry shown in Fig.~\ref{geometries}(c). The single-particle setting of Eq. \ref{H1} can be used to design a laser pulse that would generate the same exciton on a linear chain. As the exciton travels around the ring of sites, the occupation ahead of the disturbance, say at site $j+1$, is completely determined by the time-varying occupation at site $j$ because the Hamiltonian involves only nearest neighbor interactions. If the sites to the left of site 1 were hidden, for instance, the emergence of the exciton at that site and its travel to the right could be reasonably interpreted as the response to a boundary condition applied at site 1. This forms the conceptual basis for a laser-based excitation.

The approach relies on an ability to construct a chain of sites for which only the first component absorbs photons. While this can be accomplished by designing one site to have a much stronger absorption cross-section than the rest, such a strategy would imply a low exciton transport efficiency since inter-site coupling is based on the same transition dipoles as single-photon absorption.  A more promising proposition is to use a Two-Photon Absorption (TPA)~\cite{Kaiser_1961} material for the first site. The energy of individual photons, only half of the optical gap of any isolated site, would then combine to form an exciton that can be readily transferred down the chain.

The rate at which pair of photons are absorbed in TPA is given by~\cite{Rumi_2010}
\begin{equation}
\Gamma_{\rm TPA} = \frac{1.948*10^{-13}\delta I^2}{E^2_{\rm ph}} ,
\end{equation}
where $\delta$ is the TPA cross-section in Goppert-Mayer (GM) units, $I$ is the laser intensity in $W/m^2$, and $E_{\rm ph}$ is the photon energy in $eV$. There now exist organic molecules with tunable energy levels for which the TPA cross-section is on the order of $10^4$ GM~\cite{Marder_2006, Hu_2013}.  To avoid polaritonic and internal conversion influences, this rate must be substantially faster than carbon-carbon molecular vibrations, typically on the order of 50 THz. TPA absorption rates that are ten times higher than this can therefore be achieved with laser intensities in the range of $10^{10}\, \mathrm{W/cm^2}$. This is sufficiently low that the induced ponderomotive energy is less than 1 meV, so photoionization and the A.C. Stark effect are not issues. Yet another means of optically creating a localized exciton would be to use a combination of laser pulses that collectively excite an appropriate superposition of chain eigenstates~\cite{Schempp_PRL_2015}, but the TPA strategy serves to show that single-site sensitivity can be achieved.



The requisite laser pulse can be derived by comparing the laser contribution of Fig.~\ref{geometries}(b) with the periodic boundary condition of Fig.~\ref{geometries}(c). In the periodic setting, the quantum amplitudes of each site are described in the site basis, $\{\ket{j}\}_1^N$, as $u_j(t) = \braket{j|\Psi(t)}$ where $\ket{\Psi(t)}$ is the evolving state of the system for a prescribed initial condition. The eigenenergies of $\hat H$ are $\hbar\omega_j = \Delta + 2 \chi \mathrm{cos}(k_j a)$ with wavenumbers $k_j a = 2\pi j/N$. Gaussian wave packets can then be constructed as illustrated in blue in Fig. 1(c):
\begin{equation}
\ket{\Psi(0)}  = \frac{1}{\pi^{\frac{1}{4}} \sigma^{\frac{1}{2}}}  \sum_j \mathrm{e}^{\imath k_0 j a}\mathrm{e}^{\frac{-a^2(j-j_0)^2}{2 \sigma^2}} \hat{c}^{\dagger}_{j}\ket{\rm vac} .
\label{initF}
\end{equation}
Here $\sigma$ is the exciton width, the vacuum state, $\ket{\rm vac}$, is taken to be that for which all electrons reside in the valence band, $j_0$ denotes the position of the packet center, and wavenumber, $k_0$, characterizes the continuum group velocity, $v(k_0) = -2 \chi \mathrm{sin}(k_0)$. Exciton wave packet speed is estimated by tracking the position of the maximum in the packet envelope. 

The Schr\"{o}dinger equation for the ring system can be expressed as a set of N coupled ordinary differential equations for the quantum amplitudes of each ring site, $u_j$:
\begin{equation}
\imath \hbar \dot u_j = \chi u_{j-1} +  \Delta u_j + \chi u_{j+1} .
\label{ringODE}
\end{equation} 

An analogous set of equations can be constructed for the finite chain geometry of Fig. 1(b). The effect of TPA at the exposed end of the chain, combined with no resonant energy levels at the other sites, results in photon absorption only at the first site. For simplicity, the TPA dynamics are replaced by a simple light-matter interaction involving a transition dipole, $\vec\mu$, and applied electric field, $\vec E(t)$, that are taken to be parallel. The finite chain amplitudes, $q_j(t)$, then evolve according to the following equations:   
\begin{eqnarray}
\imath\hbar {{\dot q}_0} &=&  - \mu E{q_1} \nonumber \\
\imath\hbar  {{\dot q}_1} &=&  - \mu^* {E}{q_0} + \Delta{q_1} + \chi {q_2}\nonumber \\
\imath\hbar  {{\dot q}_j} &=& \chi {q_{j - 1}} +\Delta {q_j} +  \chi {q_{j + 1}}, \quad 1 < j < N  \label{chainODE} \\
\imath\hbar {{\dot q}_N} &=& \chi {q_{N  - 1}} +  \Delta{q_N}  . \nonumber 
\end{eqnarray} 
Here the ground state occupation given by
\begin{equation}
q_0(t) =  \braket{\mathrm{vac}|\Psi(t)} .
\label{groundstate}
\end{equation} 
The actual electric field is certainly real-valued, but it is useful to temporarily pretend that it is complex. A comparison of Eqs.~\ref{ringODE} and \ref{chainODE} suggests that the exciton dynamics of the ring would also be observed on the finite lattice if functions $q_0(t)$ and $E(t)$ could be achieved such that 
\begin{eqnarray}
\imath\hbar{{\dot q}_0} &=&  - \mu E{u_1} \nonumber \\
 - \mu^* {E}{q_0}  &=& \chi u_N .
\label{mustsatisfy}
\end{eqnarray}
%
%
Multiplication of the first equation by $q_0^*$ gives
\begin{equation}
q_0^* \dot{q}_0 = \imath \mu E q_1 q_0^* .
\label{rho1}
\end{equation}
Take the conjugate of Eq.~\ref{mustsatisfy}(a) and multiply by $q_0$ to obtain:
\begin{equation}
\hbar q_0 \dot{q}_0^* = -\imath \mu^* E  q_1^* q_0 .
\label{rho2}
\end{equation}
Eqs.~\ref{mustsatisfy}(b), \ref{rho1} and  \ref{rho2}  can now be combined to construct an evolution equation for the probability density of the ground state:
\begin{equation}
\hbar \dot{\rho}_0 = -2 \chi \mathrm{Im}(u_N u_1^*) .
\label{rhorate}
\end{equation}
With the initial condition that $\rho_0(0) = 1$, this can be integrated to give
\begin{equation}
\rho_0(t)  = 1 - \frac{2 \chi}{\hbar} \int_0^t \!\!\!\mathrm{d} \tau \, \mathrm{Im}(u_N(\tau) u_1^*(\tau)) .
\label{rho}
\end{equation}
The magnitude of the ground state amplitude is then $|q_0(t)| = \sqrt{\rho_0(t)} =: A(t)$.

The phase of the ground state, $\varphi(t)$, is obtained by substituting $q_0(t) = A(t) \mathrm{e}^{\imath \varphi(t)}$ into Eqs. \ref{mustsatisfy}:
\begin{eqnarray}
\imath (\dot A + \imath A \dot \varphi) &=&  - \mu E u_1 \nonumber \\
 - \mu {E}  &=& \frac{\chi u_N^* \mathrm{e}^{\imath \varphi}}{A}  .
\label{mustsatisfyphi}
\end{eqnarray}
The conjugate of the second equation can be used to eliminate $E$ from the first equation to give a rate equation for the ground state phase:
\begin{equation}
\imath \hbar(A \dot A + \imath A^2 \dot \varphi) =  \chi u_N^* u_1 .
\label{phaserate}
\end{equation}
This can be simplified by noting that $A\dot A = \frac{1}{2} \dot{\rho}_0$ and using Eq.~\ref{rhorate} to obtain:
\begin{equation}
\dot \varphi = -\frac{\chi}{\hbar\rho_0}\mathrm{Re}(u_N^* u_1) .
\label{phaseratefinal}
\end{equation}
With the initial condition of $\varphi(0)=0$, we therefore have that
\begin{equation}
\varphi (t)  = - \frac{\chi}{\hbar}  \int_0^t \!\!\!\mathrm{d} \tau \, \rho_0(\tau)\mathrm{Re}(u_N^*(\tau) u_1(\tau)).
\label{varphi}
\end{equation}
The ground state amplitude, $q_0(t) = A(t) \mathrm{e}^{\imath \varphi(t)}$, is thus completely determined. 

Eq.~\ref{mustsatisfy}(b) can then be used to construct a laser pulse that will excite an exciton on the finite lattice:
\begin{equation}
E(t) = -\frac{\chi u_N(t)}{\mu^* \sqrt{\rho(t)} \mathrm{e}^{\imath \varphi(t)}} .
\label{E1}
\end{equation}

\begin{figure}[t]\begin{center}
\includegraphics[width=0.4\textwidth]{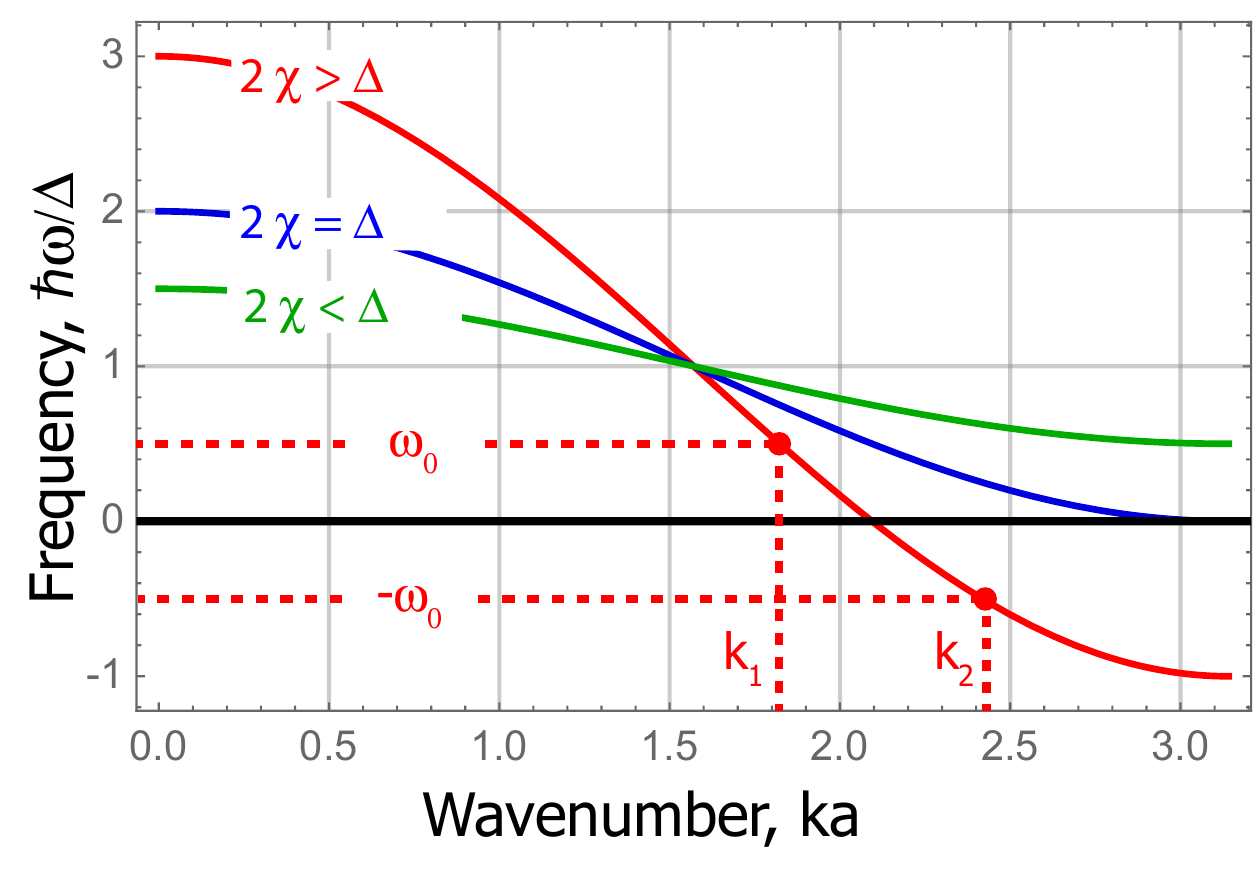}
\caption{
{\bf Dispersion Relations on a Finite Lattice}. The dispersion relation of Eq.~\ref{dispersion} is plotted for three value of hopping parameter $\chi$. When $2\chi > 1$ (red), there exist positive temporal frequencies with negative counterparts leading to the generation of multiple excitons. When $2\chi \le 1$ (blue, green), single exciton packets will be generated.}
\label{dispersionfig}
\end{center}
\end{figure}

Still holding aside its artificially complex nature, any such applied electric field must be composed of temporal frequencies that excite resonant modes of the lattice. For the finite chain of Fig. 1(b), the site basis representation of these eigenmodes is:
\begin{equation}
u_j^{(m)} = \sqrt{\frac{2}{N+1}} \mathrm{sin}(k^{(m)} j),
\label{modes}
\end{equation}
where superscript $m$ indicates the mode, $j$ is the lattice site, and the wavenumber of each mode is:
\begin{equation}
k^{(m)} =  \frac{m \pi }{N+1} .
\label{wavenumber}
\end{equation}
The associated dispersion relation,
\begin{equation}
\hbar\omega^{(m)} =  \Delta + 2 \chi \mathrm{cos}(k^{(m)}),
\label{dispersion}
\end{equation}
indicates that it is possible to have two eigenmodes of differing wavenumbers that share a common temporal rate of oscillation. This is illustrated in Fig.~\ref{dispersionfig}, where mode $k_1$ oscillates with temporal frequency $\omega_0$ while mode $k_2$ oscillates with equivalent temporal frequency $-\omega_0$. A range of such mode pairs exists provided $2\chi > 1$. Because it is the temporal oscillation of the laser that is used to excite eigenmodes, this implies that two exciton wave packets, with central wavenumbers $k_1$ and $k_2$, may result from a single laser pulse. Although it may be technologically useful to generate two energetically equivalent excitons in this way, we restrict attention to crystals for which $2\chi \le 1$ so as to produce excitons identical to those prescribed on the ring geometry.

%
%

\begin{figure}[t]\begin{center}
\includegraphics[width=0.4\textwidth]{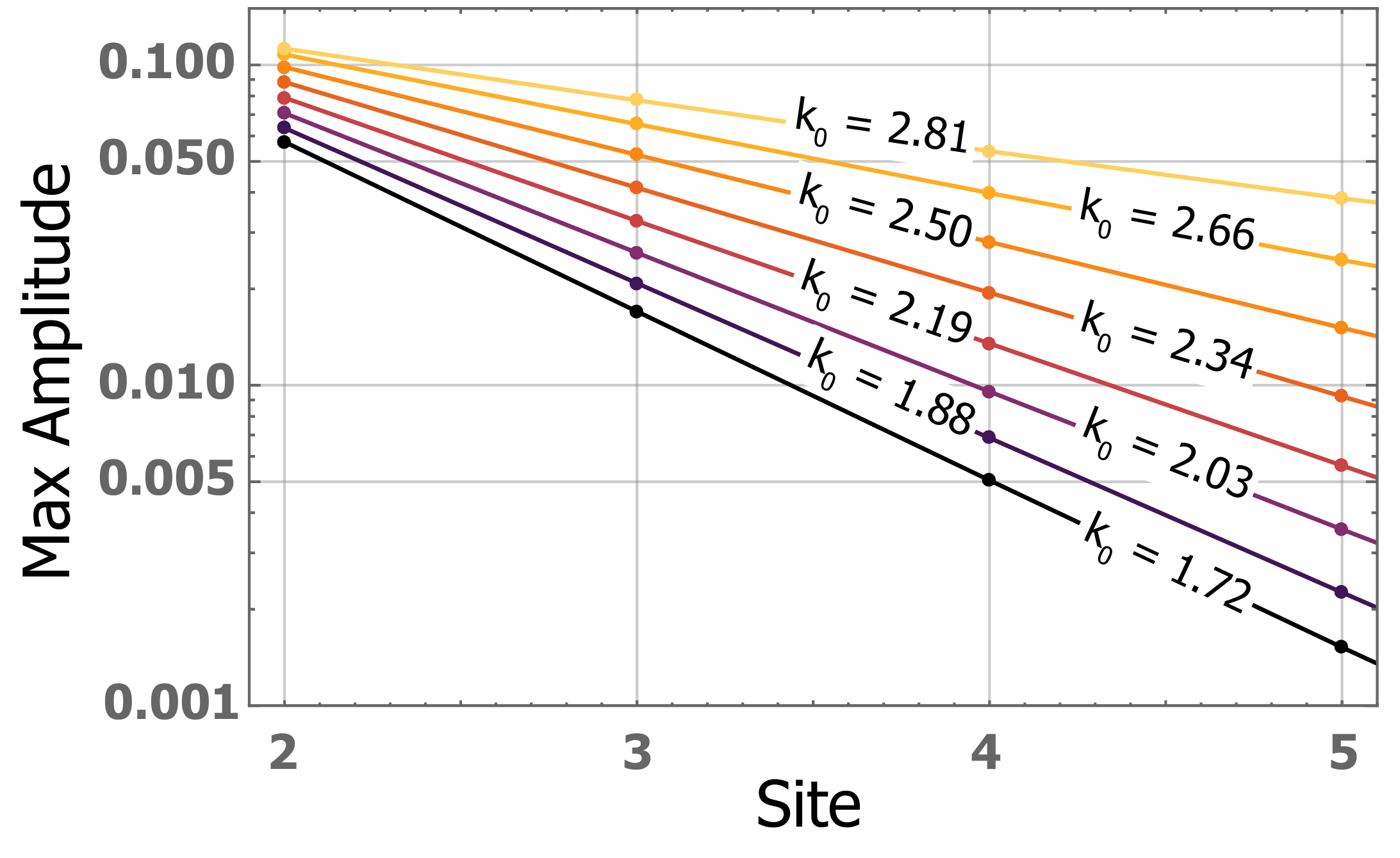}
\caption{
{\bf Quantum Interference Evanescence.} The finite lattice of Fig. 1(b) is excited with a (non-physical) laser pulse composed of only the second term, $-E*$, of Eq.\ref{decomp}. The resulting disturbance is dominated by an exponential decay of the maximum exciton occupancy that each site attains for the first few sites nearest to the excitation source. Here $N=200$, $\Delta = 2 \chi$, and $\sigma = 10.0$. The central wavenumber of the packet is labeled for each simulation set, and the straight lines are exponential fits.}
\label{QIE1}
\end{center}
\end{figure}

\begin{figure}[t]\begin{center}
\includegraphics[width=0.3\textwidth]{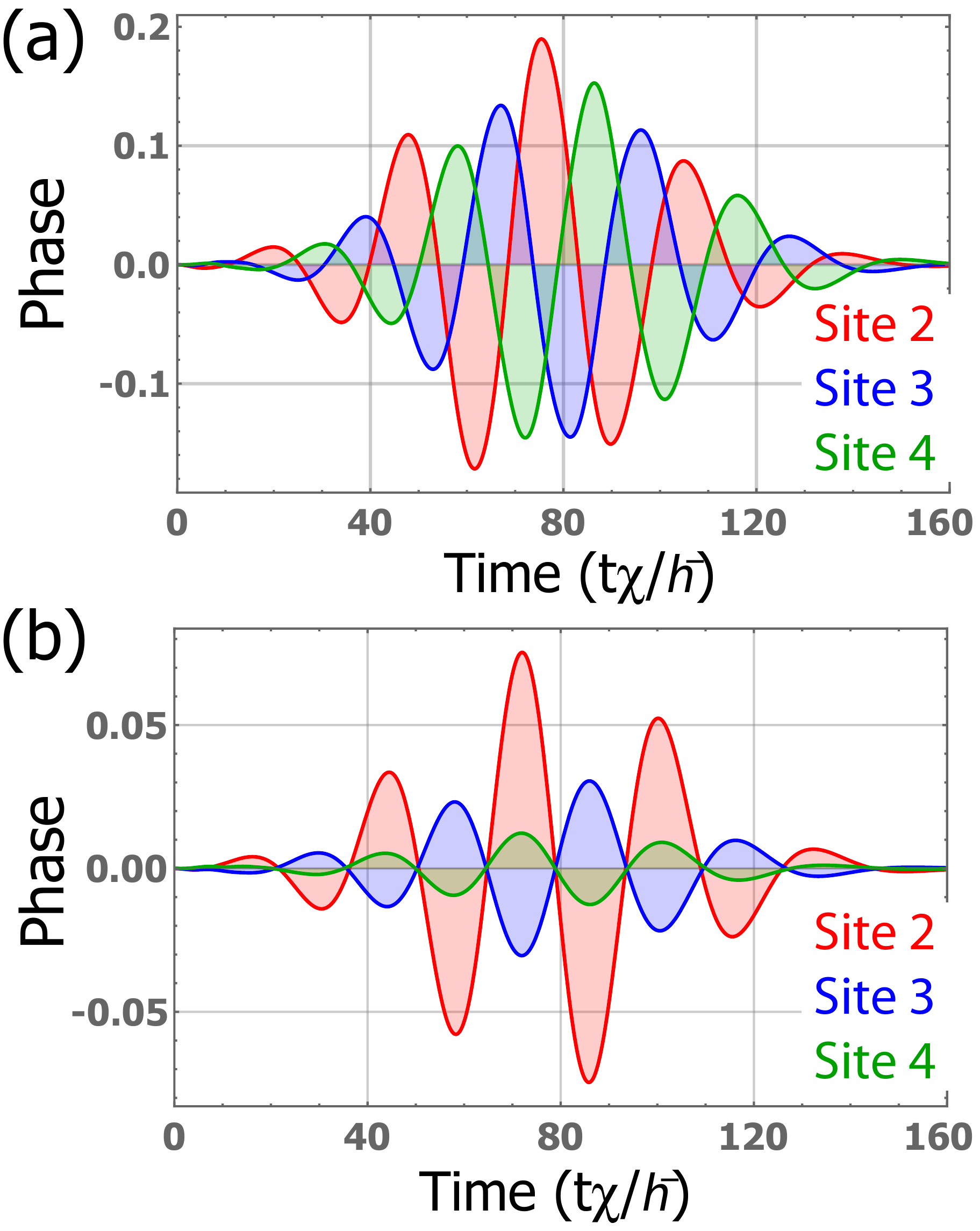}
\caption{
{\bf Quantum Interference Evanescence.} The finite lattice of Fig. 1(b) is excited with each of the terms of Eq.~\ref{decomp} separately. (a) Laser pulse consists of only the first term in Eq.~\ref{decomp}. The central wavenumber of the ring exciton is $k_0 = 2.19$ which is also the phase difference between adjacent sites since the characteristic length is the lattice spacing, $a$. (b) Laser pulse consists of only the second term in Eq.~\ref{decomp}. The amplitude of adjacent sites are out of phase by $\pi$, and this is the source of the exponential decay shown in Fig.~\ref{QIE1}.  For both plots, $N=200$, $\Delta = 2 \chi$ and $\sigma = 10.0$. }
\label{StandingWaves}
\end{center}
\end{figure}

It would at first seem that an appropriate laser pulse and parameter range has been constructed, but the form of the electric field is unphysical because it is complex valued. A path forward lies in using a simple decomposition: 
\begin{equation}
E = 2 E_{\mathrm {re}} - E^* . 
\label{decomp}
\end{equation}
The first term is real-valued and is capable of generating a wave packet that is essentially the same as that of the complex field. This is because the disturbance generated by the second term, $E^*$, is dominated by what might be referred to as \emph{quantum interference evanescence} (QIE). This QIE dies off exponentially as shown in Fig.~\ref{QIE1} for a range of central wavenumbers. The explanation for this behavior is made clear with the help of Fig.~\ref{StandingWaves}. In panel (a), only the first term in Eq.~\ref{decomp} is used to create temporal plots of the amplitude for the three sites closest to the laser pulse. For the choice of parameters listed in the figure, the phase difference between adjacent sites is $2.19$ radians. The result is a  traveling wave packet with this as its central wavenumber as will be subsequently shown. When only the second term of Eq.~\ref{decomp} is used to create an excitation, though, the resulting phase shift between adjacent sites is $\pi$ as shown graphically in panel (b) of Fig.~\ref{StandingWaves}. This generates standing waves and the excitation does not propagate. The result is consistent with the prediction of zero group velocity from Eq.~\ref{dispersion} with $k_0 = \pi$ and holds true for the parameter range of interest---i.e. $2\chi \le 1$. It should also be pointed out that the conjugate electric field does generate a tiny propagating disturbance that becomes more prominent as $\chi$ or the central wavenumber increase. This error is quantified in subsequent simulations. 

A laser pulse that can be physically implemented to generate an approximation to the desired wave packet is therefore
\begin{equation}
E_\mathrm{phys}(t) = -2\mathrm{Re}\biggl[\biggl(\frac{\chi u_N(t)}{\mu q_0(t)}\biggr)^*\biggr] .
\label{E2}
\end{equation}
%

%
%

%
%

\subsection{Real-Time, Time-Domain Density Functional Theory} 

Real-Time, Time-Domain Density Functional Theory (RT-TDDFT) simulations~\cite{Runge_1984} offer a more realistic implementation of exciton wave packet engineering. While standard DFT is a ground state theory, RT-TDDFT allows electron density to dynamically respond to laser irradiation~\cite{Peng_2015,Xavier_2015}. Inter- and intra-site electron interactions, a manifold of energy levels, and temporally varying electronic orbitals with a complex spatial character are all captured within this computational paradigm which has been successfully applied to study excited-state electron dynamics\cite{Cong2012, Yabana2012,  Lopata2011}.

The Kohn-Sham (KS) formulation of time-dependent density functional theory (RT-TDDFT) is
\begin{eqnarray}
\mathrm{i}\hbar\frac{\partial}{\partial t}\ket{\psi_i(t)} &=&\Big[ \hat T + \hat\nu_{\rm{ext}}(t) + \hat\nu_{\rm {Hartree}}[n](t) \nonumber\\
&&+ \hat\nu_{\rm{xc}}[n](t)\Big]\ket{\psi_i(t)} \label{dens},
\end{eqnarray}
where
\begin{equation}
n(r,t) = 2\sum_i^{N}  \braket{\psi_i(t)|\psi_i(t)}  
\end{equation}
is the electron density. The kets $\ket{\psi_i}$ are the time dependent Kohn-Sham (TDKS) orbitals, $\hat \nu_{\rm{ext}}$ is the external potential that accounts for the light-matter interaction, $\hat\nu_{\rm {Hartree}}$ is the Hartree potential that depends on electron density, and $\hat\nu_{\rm{xc}}$ is the exchange-correlation potential that also has a dependence on electron density. Eq. \ref{dens} gives the spin-reduced electron density and $2N$ is the total number of electrons considered.

Light-matter interaction is accounted for with a non-relativistic, semi-classical contribution to the Hamiltonian:
\begin{equation}
\hat\nu_{\rm{ext}}(t)  = \frac{1}{2 m_e}\bigl(\hat{\bf P} - e {\bf A}(t)\bigr)^2 .
\end{equation}
Here $\hat{\bf P}$ is the many-electron momentum operator and ${\bf A(t)}$ is the vector potential of the field, an explicit function of time but not position because the requisite wavelengths (hundreds of nanometers) are much longer than the dimensions of the system (nanometers). This term is incorporated into the time-dependent Runge-Gross equation~\cite{Runge_1984} where exchange and correlation effects are accounted for within a density functional formalism. Polarizability, screening, and time-dependent absorption can then be computationally measured; transition dipoles are not prescribed, and multi-electron excitations may or may not result in traveling wave packets. While vibrational effects can be included via the Hellmann-Feynman theorem~\cite{Hellmann_1937,Feynman_1939}, nuclear positions are kept fixed in the current analysis.

The associated many-body wave functions can be post-processed to estimate the time-evolving exciton populations on each site using attachment and detachment densities\cite{AD}. These can be easily derived from the time-propagated, multi-electron wavefunction:
\begin{equation}
\ket{\Psi(t)} = \ket{\psi_1(t)\cdots\psi_N(t)},
\label{TPwf}
\end{equation}
along with individual electron-hole wavefunctions defined as
\begin{equation}
\ket{\Psi_a^i (t)} = \ket{\cdots \psi_{a-1}(t) \psi_i(t) \psi_{a+1}(t) \cdots },
\label{single}
\end{equation}
Here one electron is excited from the $a^{th}$ occupied KS orbital to the $i^{th}$ unoccupied KS orbital. The time-dependent population of these single excitations are
\begin{equation}
P_a^i(t)=|\braket{\Psi_a^i|\Psi(t)}|^2,
\label{Pt}
\end{equation}
leading to the time dependent, one-particle density matrix:
\begin{equation}
\rho(t)=2\sum_i^N\ket{\psi_i(t)}\bra{\psi_i(t)}
\label{Density}
\end{equation}
The difference density, $\Delta\rho(t)$, from which attached and detached densities are derived, can now be defined:
\begin{equation}
\Delta\rho(t) = \rho(t) - \rho(0).
\label{DifDensity}
\end{equation}
This can be described as a matrix in the basis of $\{\psi_m(0)\}_{m=1}^N$, a set of KS orbitals that includes all occupied states along with a sufficient number of unoccupied states:
\begin{eqnarray}
&&(\Delta\rho(t))_{m,n} = \bra{\psi_m(0)}\Delta\rho(t)\ket{\psi_n(0)}\\
&=& 2\sum_i^N\braket{\psi_m(0)|\psi(t)}\braket{\psi(t)|\psi_n(0)}-2\delta_{mn}\nonumber.
\label{DifDensity2}
\end{eqnarray}
The eigenvalues of this matrix can be divided into those that are positive, $\{n_i^A\}$, and those that are negative, $\{n_i^D\}$, with corresponding eigenvectors $\{\phi_i^A\}$ and $\{\phi_i^D\}$. The attachment and detachment density are then defined in terms of these quantities:
\begin{eqnarray}
\rho_A&=&\sum_in_i^A\ket{\phi_i^A}\bra{\phi_i^A}\nonumber\\
\rho_D&=&-\sum_in_i^D\ket{\phi_i^D}\bra{\phi_i^D},
\label{DA}
\end{eqnarray}
Note that $n_i^A=-n_i^D$ because $n_i^A$ electrons are excited from $\ket{\phi_i^D}$ to $\ket{\phi_i^A}$. 

The attachment and detachment densities, in turn, allow the exciton population on $i^{th}$ site to be estimated as
\begin{equation}
N_X=\int_{V_i}(\rho_A+\rho_D )
\label{NofX}
\end{equation}
where $V_i$ is the effective volume associated with site $i$.

This methodology allows exciton wave packets, and in fact the more fundamental electron and hole constituents, to be tracked over time as they move down a chain of sites.  

RT-TDDFT was implemented using the Octopus code~\cite{Octopus_2006} with a Troullier-Martins pseudopotential and a Perdew, Burke, and Ernzerhof (PBE) exchange-correlation potential within the Generalized Gradient Approximation. All simulations used a time step of 0.66 as. The simulation domain was comprised of spheres created around each atom with a sphere radius 2.4 \AA\, for the benzene molecules and 4 \AA\, for methane molecules. Grid sizes of 0.15 \AA\, (benzene),  0.175 \AA\, (2-site and 3-site methane chains) and 0.175 \AA\, (20-site methane chain) were used. 

\subsection{Linking Tight-Binding and Density Functional Theory Paradigms}

In order to make quantitative comparisons between TB and RT-TDDFT predictions, the latter was used to generate the three parameters that characterize the single-particle TB Hamiltonian of Eq. \ref{H1}: site energy, $\Delta$, nearest neighbor coupling, $\chi$, and transition dipole, $\mu$. These are nontrivial procedures that are explained below.

\subsubsection{Site Energy, $\Delta$}

Tight-binding calculations are based on a knowledge of the site-centered, diabatic, excitonic energy associated with a chain of methane molecules, $\Delta$ of Eq. {H1}. For weakly coupled sites, this can be estimated perturbatively provided the coupling energy is known. It is also possible to estimate this nonperturbatively using Fragment Energy Differencing (FED)~\cite{FED1, FED2, FED3} or Edmison-Ruedenberg (ER) localization~\cite{Subotnik2, ER}. These rely on static DFT analysis, but it is possible to exploit our explicit-time setting to estimate $\Delta$ using RT-TDDFT. 

\begin{figure}[t]\begin{center}
\includegraphics[width=0.45\textwidth]{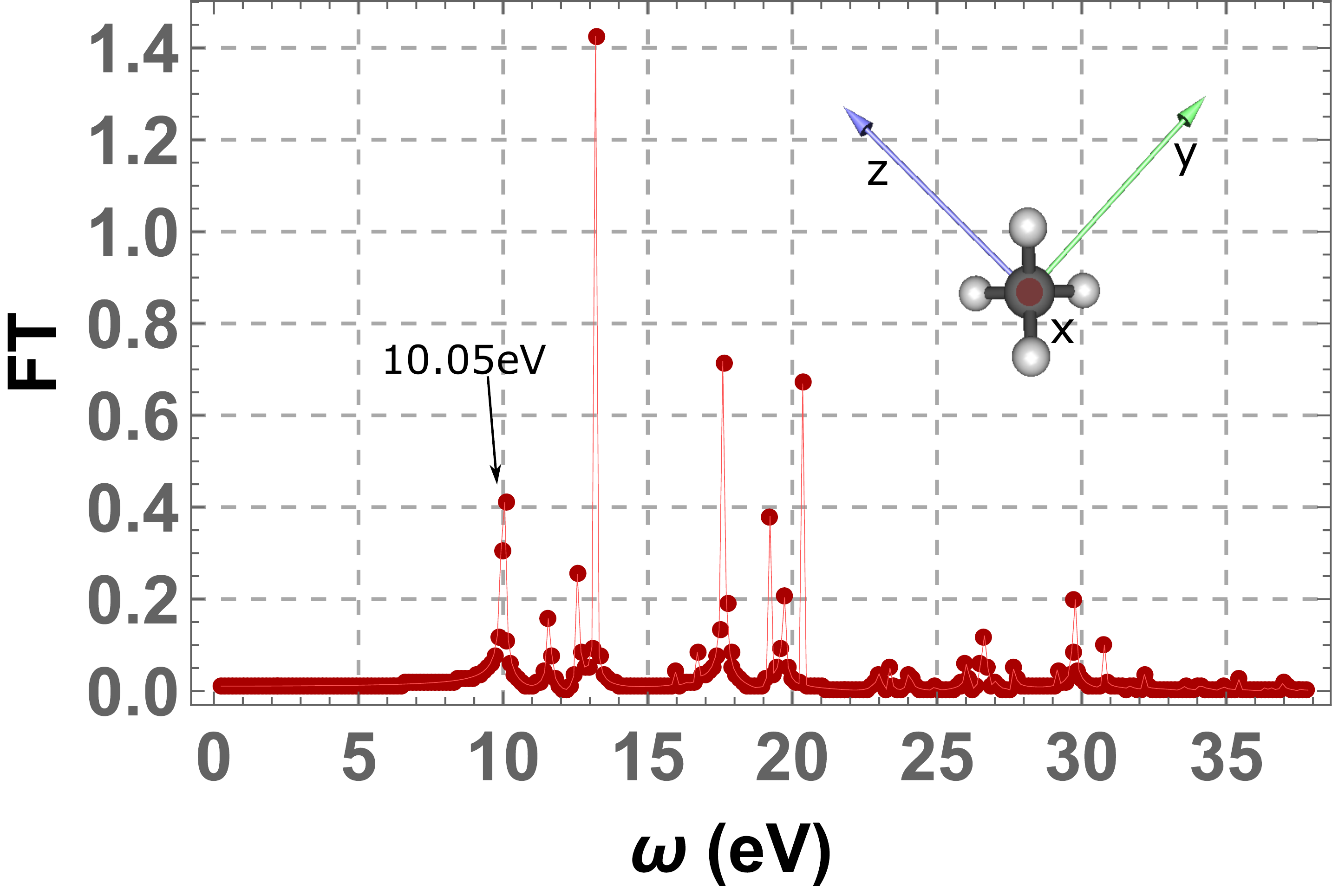}
\caption{
{\bf Fourier Transform of Dipole Moment for a Single Methane Molecule.} A Discrete Fourier Transform (FT) of the dipole moment versus time, after a delta kick polarized in x direction, is used to determine the lowest excitation energy, 10.05 eV. Discrete points are the actual Fourier data while the red lines are a guide to the eye.}
\label{FT}
\end{center}
\end{figure}

This new methodology begins by identifying the lowest exciton energy for a single methane. The molecule is excited by a laser impulse,
\begin{equation}
E_x(t)=A \delta(t)\hat{x},
\label{kick}
\end{equation}
where A is the strength of an impulsive kick to the system, and the three-fold molecular degeneracy of the lowest exciton state is broken by orienting the laser along the x-axis as shown in the inset of Fig.~\ref{FT}. The radiation absorbed over 33 fs generates a time-varying dipole moment,
\begin{equation}
\mu_x(t)=\bra{\Psi(t)}\hat X\ket{\Psi(t)},
\label{Dipole}
\end{equation}
with the spectral profile shown in Fig.~\ref{FT}. As indicated in the plot, the lowest (polarized) excitation energy of an isolated methane molecule is 10.05 eV.  

%

\begin{figure}[t]\begin{center}
\includegraphics[width=0.45\textwidth]{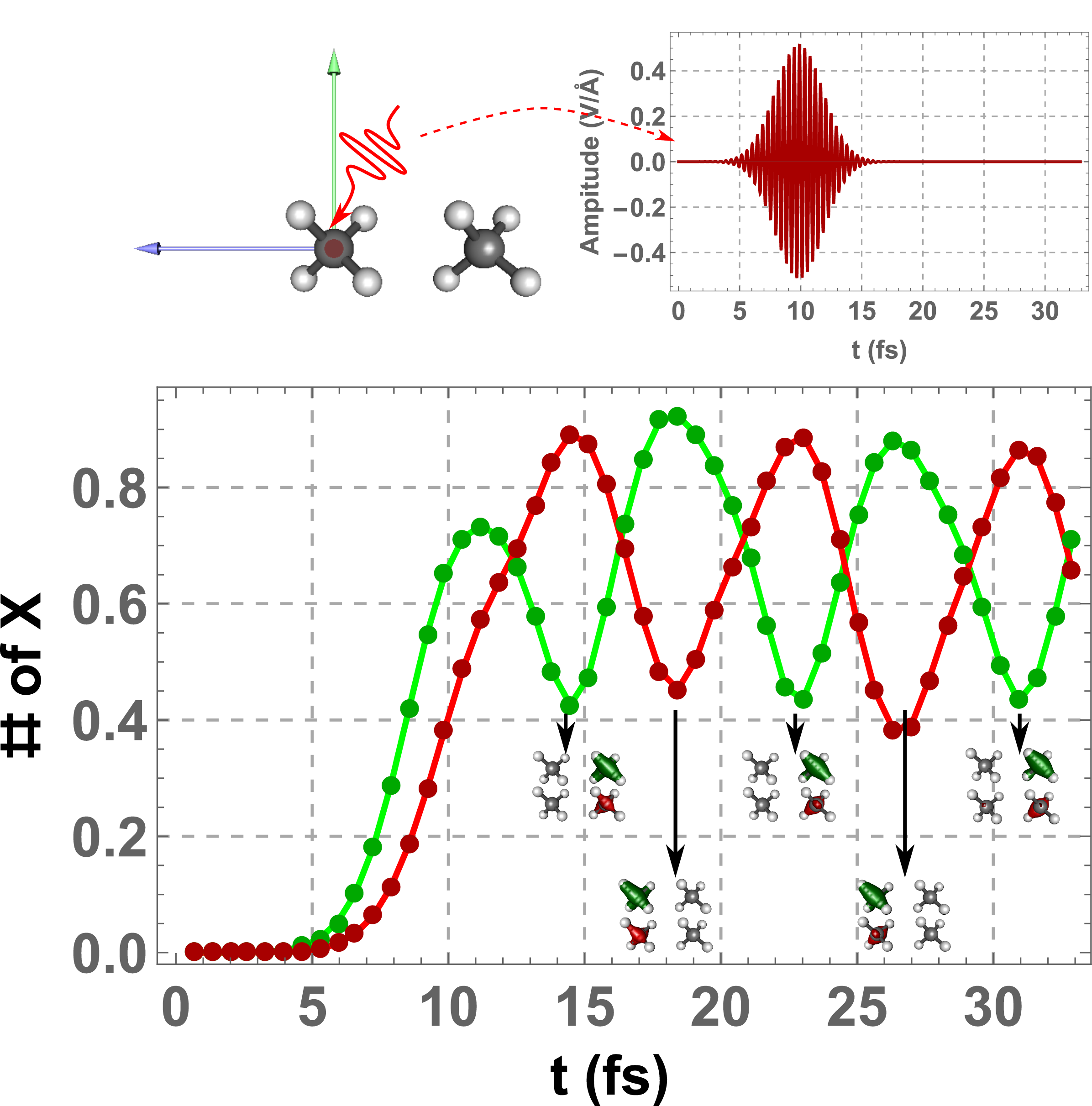}
\caption{
{\bf Rabi Oscillation in Methane Dimer.} A set of gaussian-shaped laser pulses, Eq. \ref{laser}, were applied to the left site of methane dimer to determine the energy best able to create a clean exciton oscillation. The laser was polarized along x direction with $F=0.5\, \mathrm{V}/\mathrm{\AA} $, $\tau=1.97\, fs$, and $t_0=9.87\, fs$. The optimum value of $\hbar\omega$ was found to be $9.6$ eV and the associated oscillations are shown. The red (green) line is the exciton occupation on the right (left) site. The isosurfaces of electron (green) and hole (red) density at $0.008 /{\rm Bohr}^3$ are shown underneath these curves. }
\label{CH4dimer}
\end{center}
\end{figure}

This energy was used as a starting point to quantify the site-centered, diabatic, excitonic energy associated with a chain. To account for the influence of neighboring sites, it was deemed sufficient to consider a two-site system with the left molecule excited and the right molecule serving as a proxy for the remainder of the chain. A gaussian laser pulse was applied on the left site of the dimer:
\begin{equation}
\vec E(t) = F \,{\rm cos}(\omega t) {\rm exp}\biggl(\frac{-(t - t_0)^2}{2 \tau^2}\biggr).
\label{laser}
\end{equation}
A range of excitation energies, $\hbar \omega$, were considered around 10.05 eV to determine the value that gives a clean (single frequency) Rabi oscillation between the sites. As shown in Figure~\ref{CH4dimer}, this occurs for $\hbar \omega = 9.6$ eV, and it was this value of diabatic exciton energy that was used in our TB analysis for a direct comparison with RT-TDDFT results.

%
%

\subsubsection{Coupling Between Nearest Sites, $\chi$}

\begin{figure}[t]\begin{center}
\includegraphics[width=0.45\textwidth]{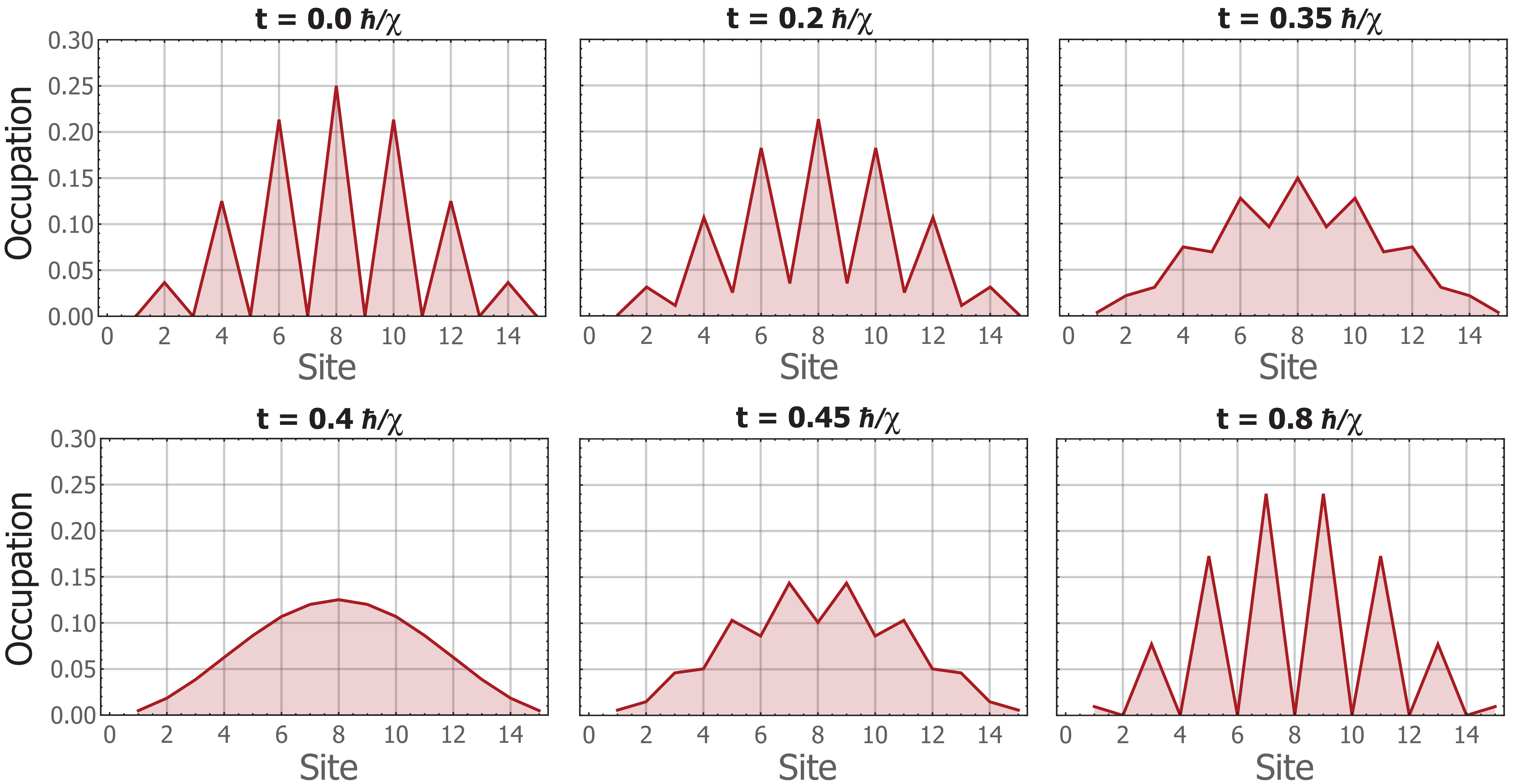}
\caption{
{\bf Oscillation in an N-Site Chain.} The initial condition of Eq. \ref{init} was applied to a 15-site chain to generate the oscillation pattern shown. }
\label{Nsite}
\end{center}
\end{figure}

The TB coupling parameter, $\chi$, can be derived from RT-TDDFT analysis by measuring the rate of oscillation in site populations due to a prescribed laser pulse. In the simplest case, a Rabi oscillation can be established in a two-site system, but oscillations associated with multiple sites are possible as well. This is relevant since the coupling between two isolated sites, each with just one nearest neighbor, is different from the coupling between sites with neighbors to both right and left. To this end, we considered the TB Hamiltonian for a chain of sites with identical site energies and hopping parameters:
\begin{equation}
\hat{H}  = \frac{\Delta}{2}\sum_{j}\hat{n}_{j} + \chi\!\!\!\!\!\!\sum_{<i,j>,i\neq j}\hat{c}^{\dagger}_{j}\hat{c}_{i}  + \rm{H.c.}
\label{Hamiltonian}
\end{equation}
Let $\ket{\varphi_j}$ and $\varepsilon_j$ be the associated eigenkets and eigenvalues, respectively. Also define an auxiliary operator, 
\begin{equation}
\hat{\Gamma}  = \sum_{<i,j>,i\neq j}\hat{c}^{\dagger}_{j}\hat{c}_{i}  + \rm{H.c.}
\label{Auxiliary}
\end{equation}
and denote the difference between its maximum and minimum eigenvalues as $\gamma$, a function of number of sites. Now prepare the initial state of the system:
\begin{equation}
\ket{\Psi_{\rm{init}}} = {\frac{1}{\sqrt{2}}\biggl(\ket{\varphi_{{\rm max}}} - \ket{\varphi_{{\rm min}}}}\biggr),
\label{init}
\end{equation}
Here the two eigenkets are associated with the maximum and minimum eigenvalues of $\Gamma$, respectively. The ensuing dynamics will then exhibit an oscillation in the site populations, as shown in Fig. \ref{Nsite}, in which the odd-numbered sites have a population that oscillates. The relationship between coupling, $\chi$, and the oscillation period, T, is easily derived to be
\begin{equation}
\chi=\frac{2\pi \hbar}{\gamma \,T} .
\label{coupling}
\end{equation}
%

%

Eq. \ref{coupling} establishes the algorithm with which TB coupling can be measured in the RT-TDDFT setting for any number of sites. In the present case, since interactions beyond nearest neighbor methane molecules are very small, it is sufficient to estimate the coupling using just three sites. 

\begin{figure}[t]\begin{center}
\includegraphics[width=0.45\textwidth]{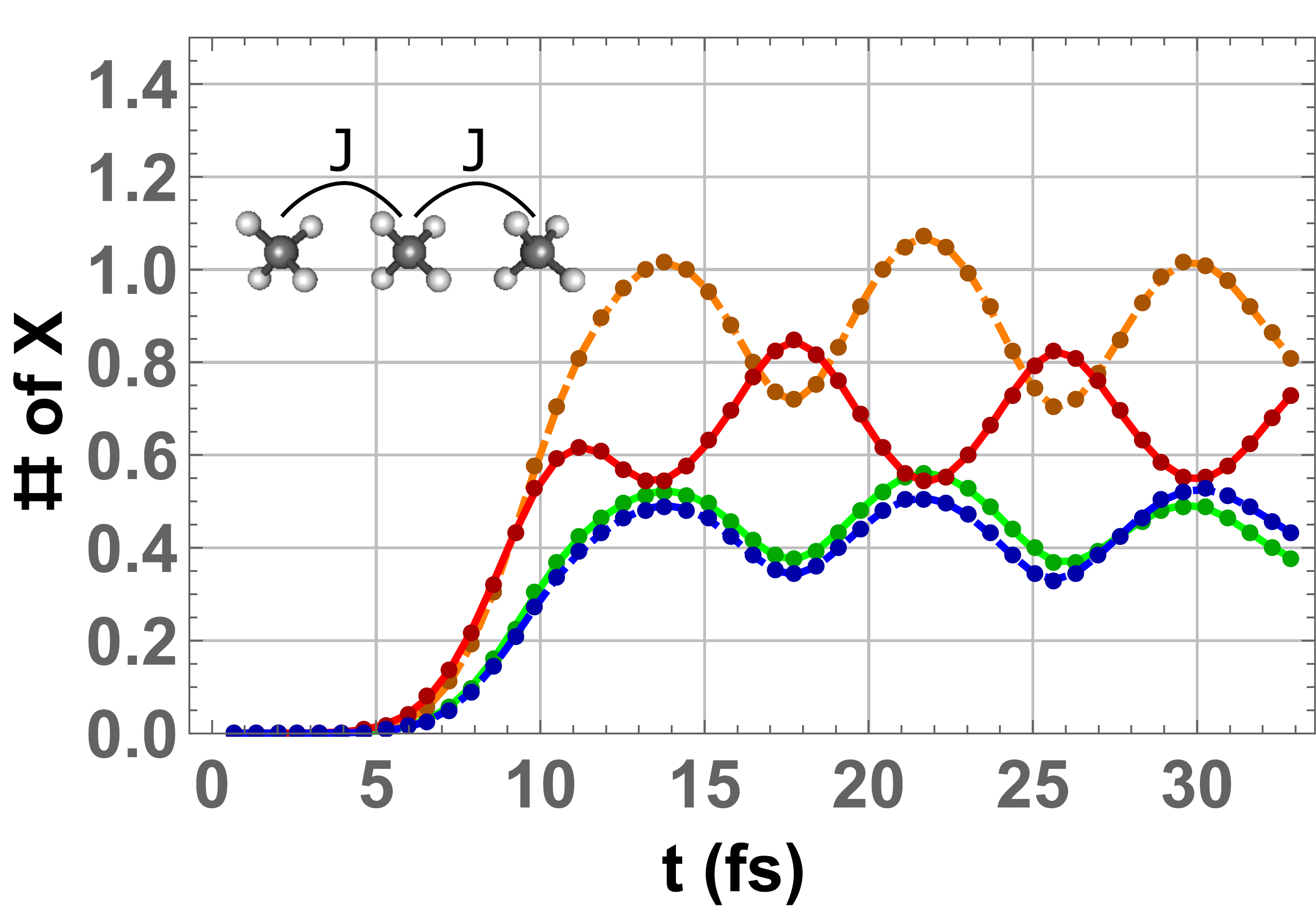}
\caption{
{\bf Oscillation in Trimer.} A gaussian laser pulse polarized along x direction, $E_x(t)=F\cos{(\omega t)}\mathrm{e}^{\frac{-(t-t_0)^2}{2\tau^2}}$ with strength $F = 0.5\, V/\AA $, $\hbar \omega = 9.6$ eV, $\tau=1.97$ fs, and $t_0 = 9.87$ fs, is applied on middle site of the trimer. Oscillation between even and odd sites results. Red indicates the exciton number on middle site,  orange is exciton number on either side, green denotes the exciton number on left site, and blue line is exciton number on right site. }
\label{CH4trimer}
\end{center}
\end{figure}

A gaussian laser pulse applied to the left site of the dimer system of Fig.~\ref{CH4dimer} approximates the initial condition of  Eq. \ref{init} and results in oscillations with a period of 8.12 fs. An analogous excitation of the center site of the trimer system of Fig.~\ref{CH4trimer}, on the other hand, gives an oscillation period of 8.23 fs. Using Eq.~\ref{coupling}, these periods correspond to coupling value of $\chi = 0.25$ eV (dimer) and $\chi = 0.18$ eV (trimer). This reflects the fact that the two-site system has a different electronic structure between sites that does the trimer system. The trimer coupling, with neighbor interactions to either side, is the one used in the 20-site simulation because it more accurately reflects the nearest-neighbor interactions of multi-site chain.
%
%

\subsubsection{Transition Dipole, $\mu$}
We also need to be able to generate a transition dipole strength directly from RT-TDDFT so that it can be used in comparative TB simulations. To make this link, we used RT-TDDFT  to apply a laser to the first site of a methane dimer to put it into its lowest excited state. The transition dipole is then defined as
\begin{eqnarray}
\mu(t)&=&\bra{\Psi(0)}\sum_i^Nr_i\ket{\Psi(t)}\nonumber\\
&\approx&\sum_i n_i \bra{\psi_i^D}r\ket{\psi_i^A}.
\label{TransDipole}
\end{eqnarray}
As shown in Fig.~\ref{CH4dimer} for $t = 18.424$ fs, the exciton is localized on left site and with a total of 0.69 electron-hole pairs. The transition dipole at $t=18.424$ fs for one exciton is then calculated to be $-0.014 - \imath 0.038\, q\cdot \mathrm{\AA}$, where $q$ is the charge of one electron. This is the value used in the comparative TB analysis.

\section{Tight-Binding Results}

\begin{figure}[t]\begin{center}
\includegraphics[width=0.45\textwidth]{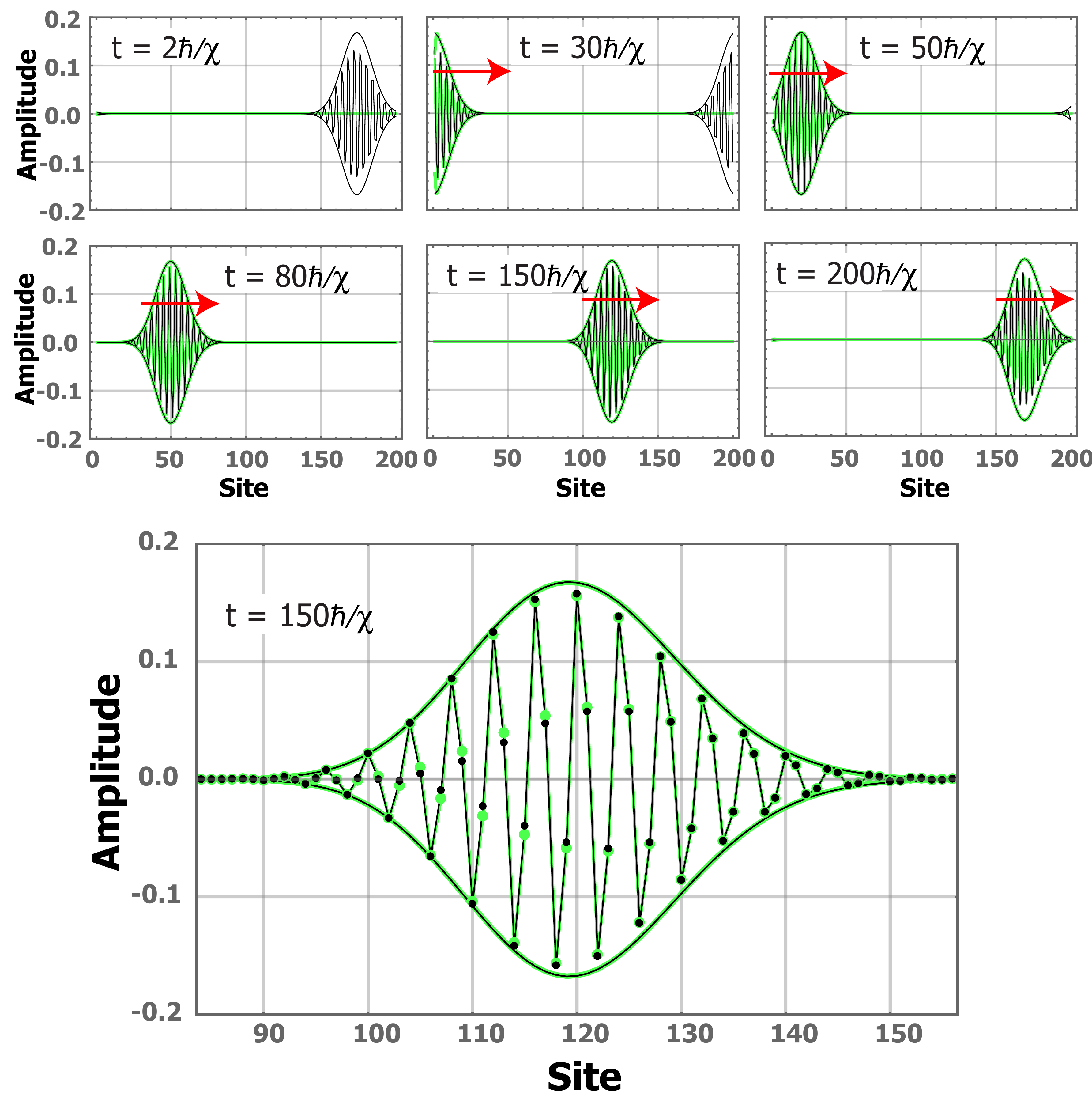}
\caption{
{\bf Evolution of a Laser-Induced Exciton Wave Packet.} The finite lattice of Fig.~\ref{geometries}(b) is excited with a laser pulse at left. The result is an exciton (green) that is plotted for several time slices along with the original exciton (black) of Fig.~\ref{geometries}(c). Here $N=200$, site energy $\Delta = 2\chi$ and the exciton footprint has a standard deviation $\sigma = 10.0a$. The central wavenumber of the ring exciton is $k_0 = 1.58/a$. The initial exciton occupation fraction of the ring is 0.5, although this can be set to any value.}
\label{Packet_Evolution}
\end{center}
\end{figure}

The proposed laser pulse methodology can now be tested by comparing excitons prescribed on the ring to those generated by on a chain via an engineered pulse of radiation. A representative result is shown in Fig.~\ref{Packet_Evolution}. The exciton for the ring (black) and that from the laser (green) are essentially indistinguishable for all time slices shown. A magnified view of one time slice (bottom panel) is required to show that any difference exists. It can be quantified with a dimensionless RMS error, by taking the difference between the two amplitudes, $\delta_j(t)$:
\begin{equation}
\epsilon_{\mathrm{rms}}(t) = \sqrt{\frac{1}{N}\sum_{j=1}^N |\delta_j(t)|^2}.
\label{rms}
\end{equation}
The RMS error at the time shown in the bottom panel is $\epsilon_{\mathrm{rms}} = 0.0042$.

%
%

\begin{figure}[t]\begin{center}
\includegraphics[width=0.3\textwidth]{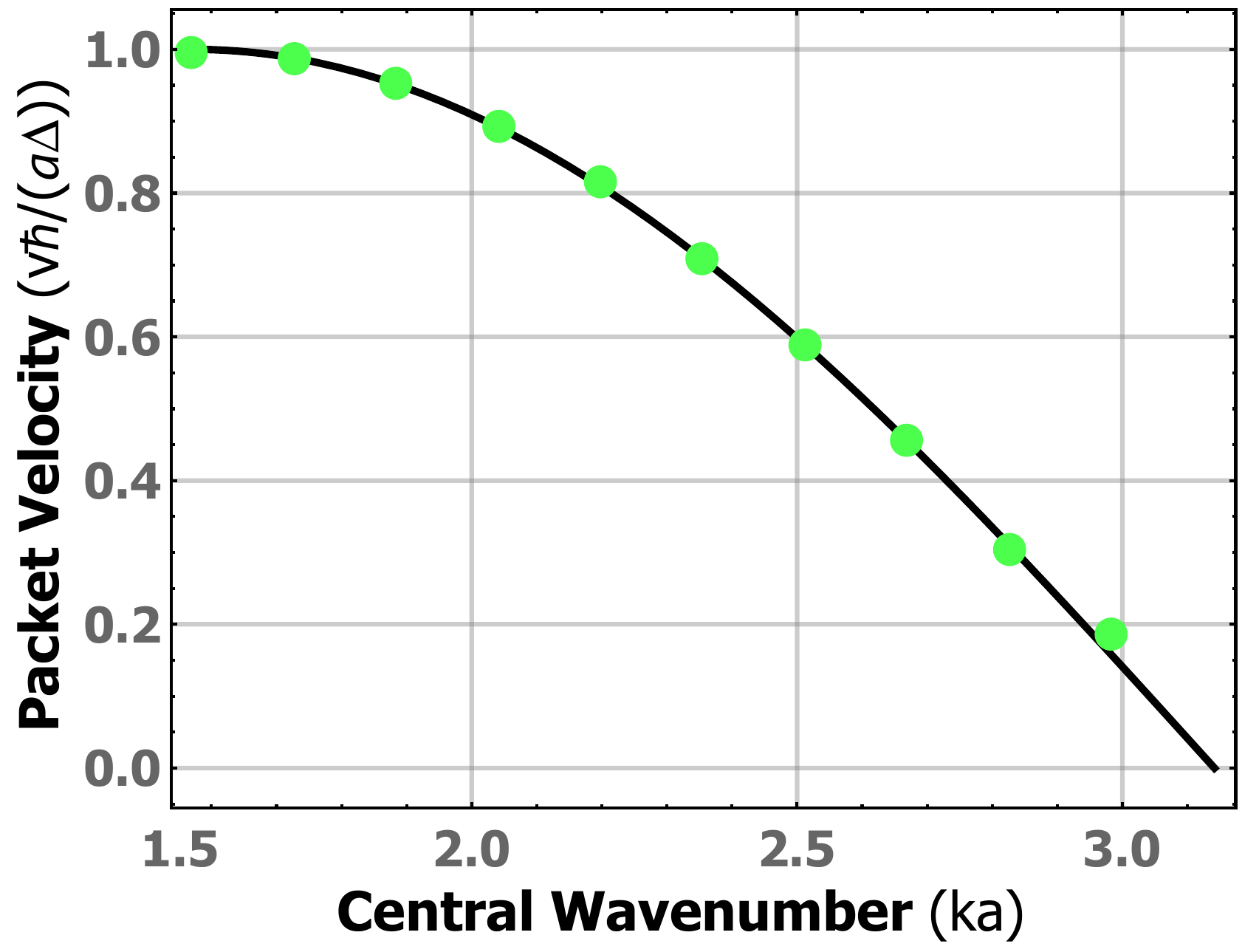}
\caption{
{\bf Tunable Exciton Speed.} The finite lattice of Fig.~\ref{geometries}(b) is excited with twice the real part of the laser pulse of Eq.~\ref{E1} for a range of central wavenumbers. The numerically measured exciton speeds (green) are compared with the speed predicted from the continuum dispersion relation (black). The speed was changed by over a factor of five in the simulations carried out. Here $N = 200$, $\Delta = 2\chi$  and $\sigma = 10.0a$.} 
\label{packetspeed}
\end{center}
\end{figure}

A series of simulations was generated in this way to numerically measure the exciton speed for a range of laser pulses. The results, plotted in Fig. \ref{packetspeed}, show that the speeds correspond to those predicted from the continuum dispersion relation of Eq. \ref{dispersion}. Significantly, it is possible to change the exciton speed by more than a factor of five through appropriate tailoring of the laser pulse. 

%
%

\begin{figure}[t]\begin{center}
\includegraphics[width=0.45\textwidth]{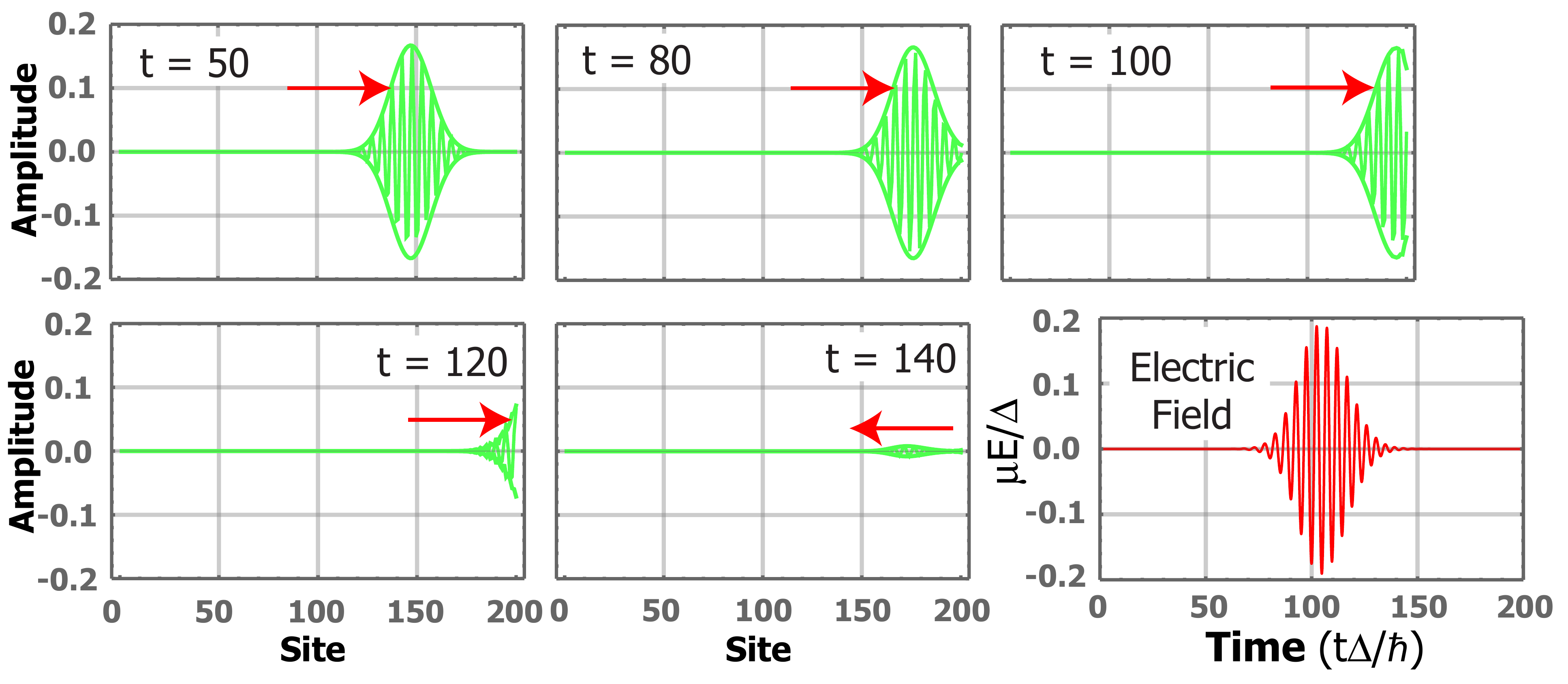}
\caption{
{\bf Exciton Annihilation.} A exciton wave packet (green) travels to the right on the finite lattice of Fig. 1(b). The same exciton (black) is considered on the superimposed ring geometry of Fig. 1(c). This is used with Eq.~\ref{E4} to design a laser pulse that will extract the excitonic energy from the right-most site. Several time slices show that the resulting stimulated emission removes the excitonic wave packet. Here $N = 200$, $\Delta = 2\chi$  and $\sigma = 10.0a$. The central wavenumber of the ring exciton is $k_0 a = 2.19$.}
\label{Annihilation}
\end{center}
\end{figure}

Just as it is possible to create excitonic wave packets, a laser field can be used to remove them as well---a synchronized version of stimulated emission. To examine this, suppose that an exciton is traveling to the right as shown in the upper left panel of Fig.~\ref{Annihilation}. In this setting, the right-most site is assumed to have a transition dipole that is perpendicular to the rest so that it can be illuminated in isolation by an applied electric field. An analysis analogous to that used to produce Eqs. 10, 14 and 15 then delivers the requisite laser pulse:
\begin{equation}
E(t) = -\biggl(\frac{\chi u_1(t)}{\mu q_0(t)}\biggr)^* .
\label{E4}
\end{equation}
Here the $|q_0(t)| = \sqrt{\rho_0(t)}$ with ground state density, $\rho_0(t)$, given by
\begin{equation}
\dot{\rho}_0 = -2 \chi \mathrm{Im}(u_1 u_N^*) .
\label{rhorate2}
\end{equation}
The evolving phase of the ground state is 
\begin{equation}
\dot \varphi = -\frac{\chi}{\rho_0}\mathrm{Re}(u_1^* u_N) .
\label{phaseratefinal2}
\end{equation}
To produce Fig.~\ref{Annihilation}, the initial ground state probability density was taken to be $0.5$ with an initial ground state phase of zero. Note that a tiny reflected excitonic packet is produced as part of the annihilation event. The associated RMS error is $\epsilon_{\mathrm{rms}} = 0.0022$ at the final time step. 
%
%

\begin{figure}[t]\begin{center}
\includegraphics[width=0.45\textwidth]{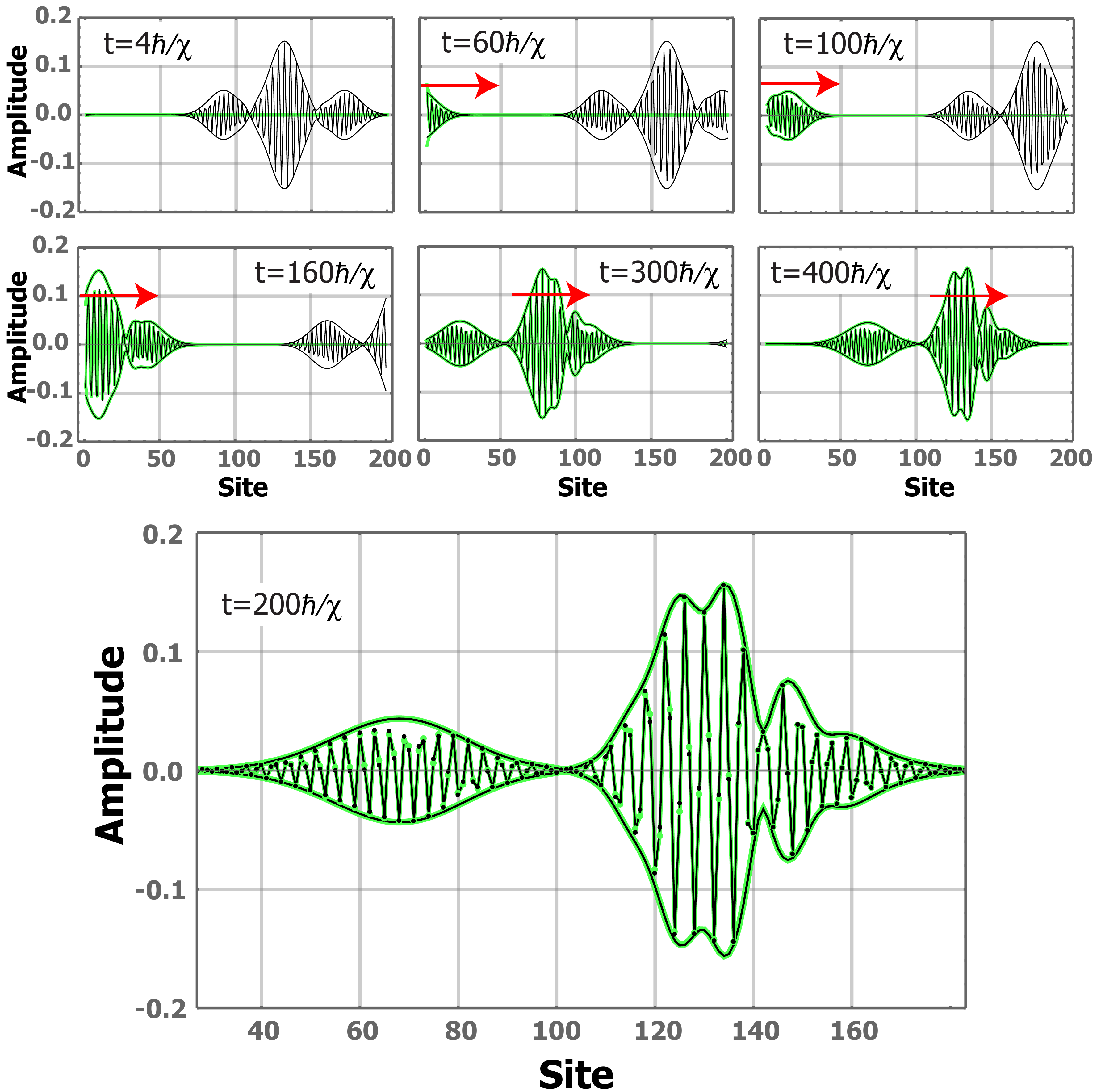}
\caption{
{\bf Control Over Wave Packet Character.} The finite lattice of Fig. 1(b) is excited with a laser pulse at left defined by Eq. 15. This creates a triplet of overlapping Gaussian excitations (green) plotted for several time slices along with the original ring wave packet (black) of Fig. 1(c). The central packet was intentionally tuned so that it travels slightly faster than its neighbors.  Here $N = 200$, $\Delta = 2\chi$  and $\sigma = 10.0a$.  The central wavenumbers of the ring exciton are, from left to right, $k_0 a= 2.03$, $k_0 a = 1.59$, and $k_0 a = 2.03$. Units of time are $\hbar/\chi$.}
\label{Packet_Shape}
\end{center}
\end{figure}

The methodology developed also allows for more complex excitations. Any envelope desired can be input as an initial condition within the ring setting with Eq. \ref{E2} then used to engineering the requisite laser pulse. For example, wave packets with a non-Gaussian profile and composed of multiple bands of wavenumbers can be produced as shown in Fig.~\ref{Packet_Shape}. 

%
%

\section{RT-TDDFT Results}

Two types of molecular chains were used to create and examine exciton wave packets in the many-body, multi-level setting offered by RT-TDDFT. In both cases, a sensitizing TPA molecule was not considered explicitly and, instead, the molecular chains were idealized so that irradiation only occurs at the first (left-most) site. The first analysis, using a short benzene chain, simply demonstrates that it is possible to generate an exciton packet with a well-defined speed. Only five sites (molecules) are considered because the associated calculations are extremely demanding. This is also the motivation for using such small molecules, and the high excitation energy would preclude their use in a physical implementation. The intent here is only to provide a computationally tractable proof of concept. 

The second implementation extends the number of sites to twenty but required that we further reduce the size of each molecule to methane. This chain is sufficiently long to quantify the speed of excitons for a range of engineered laser pulses.

\subsubsection{5-Site Benzene Chain}

\begin{figure}[hptb]
\begin{center}
\includegraphics[width=0.45\textwidth]{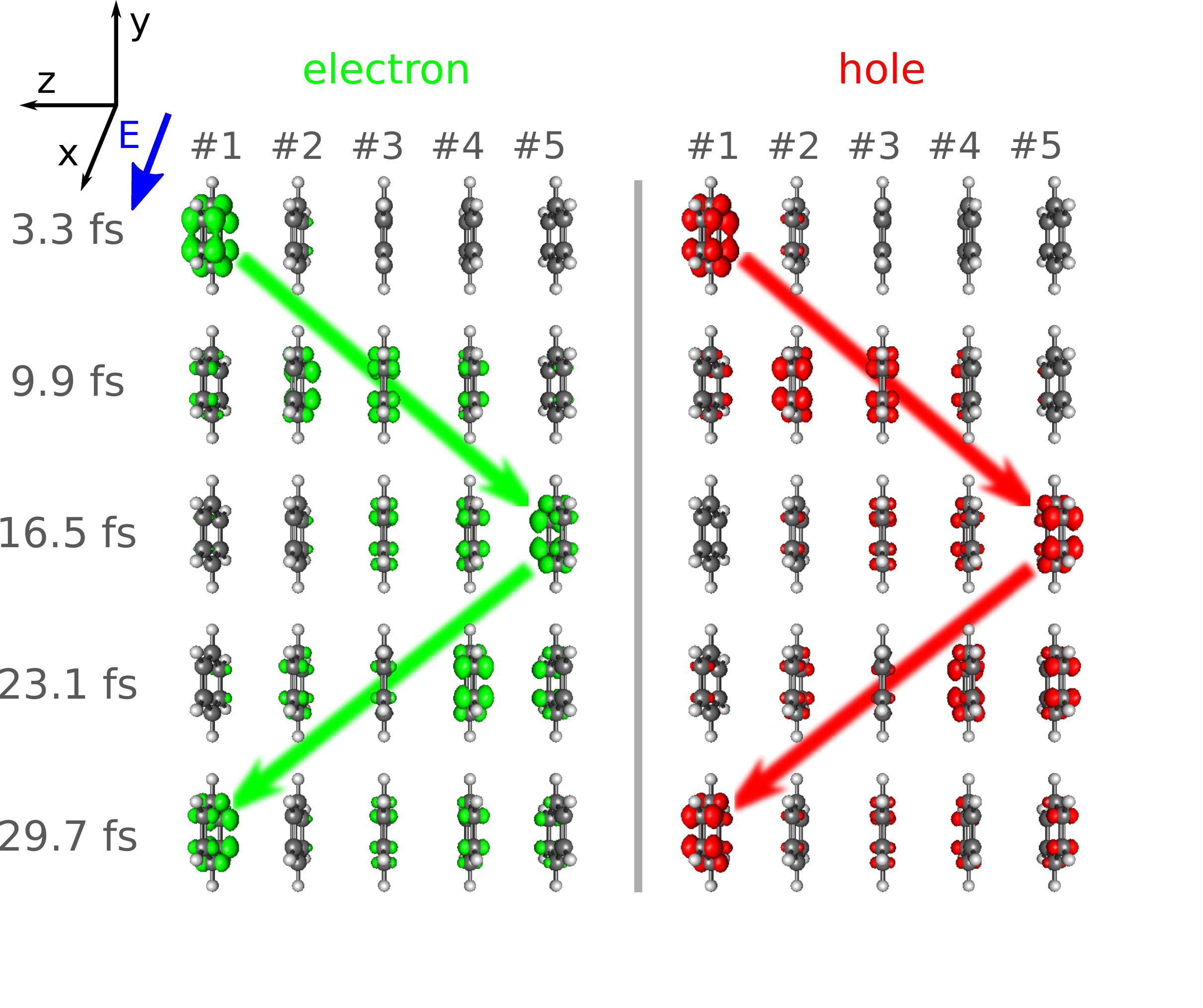}
\end{center}
\caption{ {\bf5-Site Benzene Chain.} Uncontrolled laser-generated exciton wave packet moves down a chain of 5 co-facial benzene molecules. The green and red are the isosurfaces with value 0.002 electrons per $\mathrm{Bohr}^{3}$  for electron and hole densities respectively. }\label{BenChain}
\end{figure}

An RT-TDDFT analysis was first performed on a 5-site co-facial benzene molecular chain, as shown in Fig.~\ref{BenChain}, with a 7.56 Bohr separation between sites. No attempt was made to engineer the laser pulse so as to control the shape and speed of the resulting exciton packet.  Instead, a simple Gaussian shape enveloped was applied to the first molecule: $E_0\cos(\hbar \omega t) e^{\frac{-(t-t_0)^2}{2\tau^2}}$. Its energy, $\hbar\omega=5.39$ eV, corresponds to the lowest excitation energy of a single benzene molecule. A field strength $E_0 = 0.51\, \mathrm{V}/\mathrm{\AA}$, peak time, $t_0=3.29$ fs and envelope width $\tau=0.66$ fs were used.  As is clear from Fig.~\ref{BenChain}, it is possible to generate a rudimentary exciton packet that moves down the chain, reflects at the end, and then propagates back to the left. 

%
%

\subsubsection{20-Site Methane Chain}

The second example, and the setting of our primary computational results, considers a 20-site chain of methane molecules for which the orientation, spacing and laser intensity were carefully engineered.  TB parameters, distilled from the RT-TDDFT setting as described earlier, were used to design a set of laser pulses that would generate excitons with a range of speeds. These laser pulses were then applied to both TB and RT-TDDFT paradigms, and the speeds of the wave packets generated were then estimated by linearly fitting their peak amplitudes. 

\begin{figure}[hptb]
\begin{center}
\includegraphics[width=0.45\textwidth]{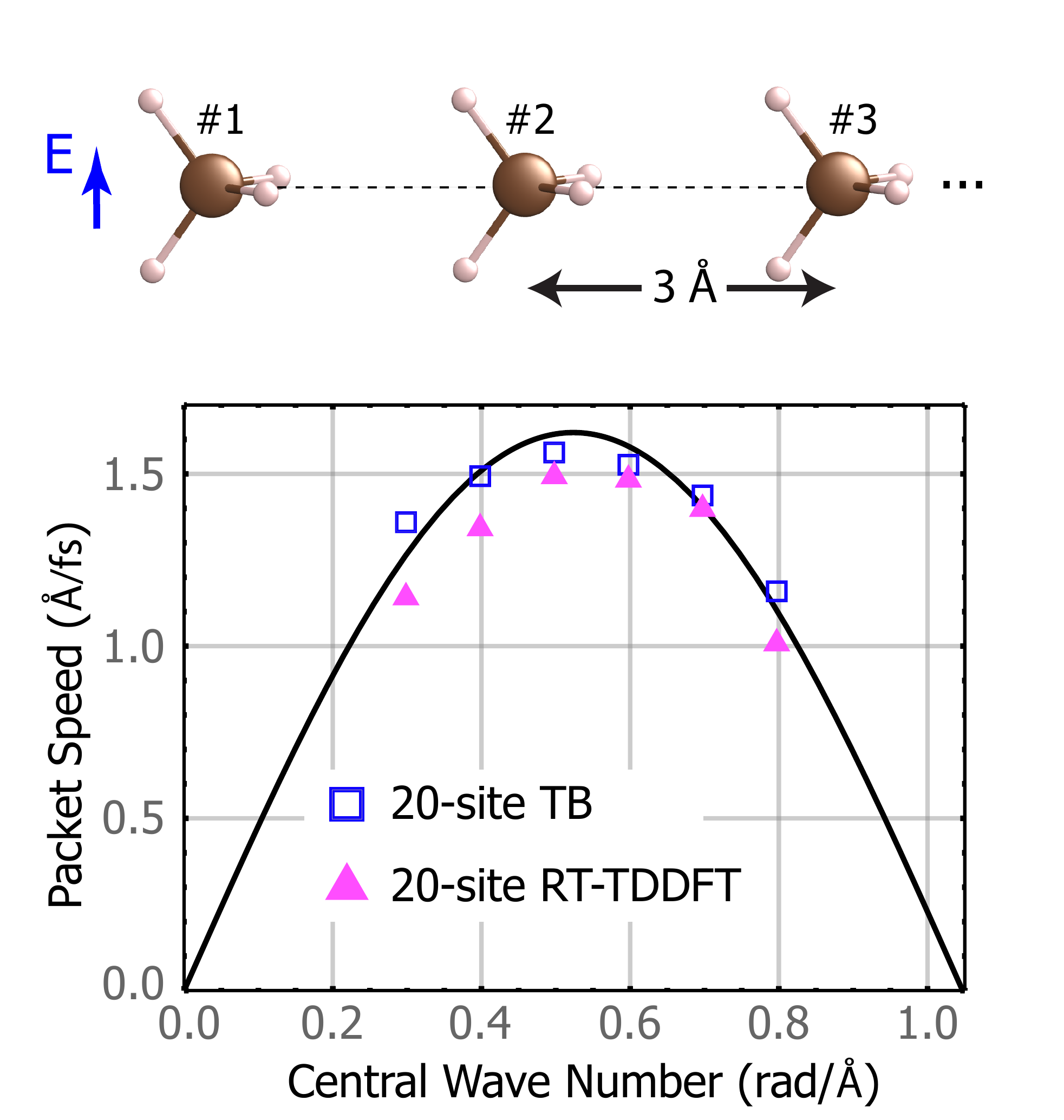}
\end{center}
\caption{{\bf 20-Site Methane Chain.} Laser pulse shape and composition is modified so as to vary exciton speed on a 20-site chain of methane molecules. Solid curve is from dispersion relation for an infinite chain (Eq. \ref{dispersion}), hollow blue squares are from 20-site TB model, and filled magenta triangles are from RT-TDDFT simulations.}
\label{Methane_Chain}
\end{figure}

The results are shown in Fig. \ref{Methane_Chain} along with the analytical group velocity obtained from the dispersion relation of Eq. \ref{dispersion}. As expected, the TB speeds are in accord with the analytical group velocity. However, the RT-TDDFT speeds are also quite close to theory. This is remarkable since each molecule offers a complex many-body electronic environment. The results demonstrate that tunable excitons can be generated in physical systems and offers a starting point for the consideration of more realistic atomic settings in which such a correspondence with TB is not possible. Exciton engineering can be carried out using only RT-TDDFT and periodic domains of any dimension.

%
%
%

\section{Exciton Manipulation}
With a methodology now in place for the laser-creation of engineered exciton wave packets, attention is turned to ways in which they can be manipulated. For a range of excitonic binding energies, a single-particle TB idealization is reasonable and the only new physics is that it is the quantum interference of excitons, and not electrons or photons, that is being used as a handle. 

\begin{figure}[t]\begin{center}
\includegraphics[width=0.45\textwidth]{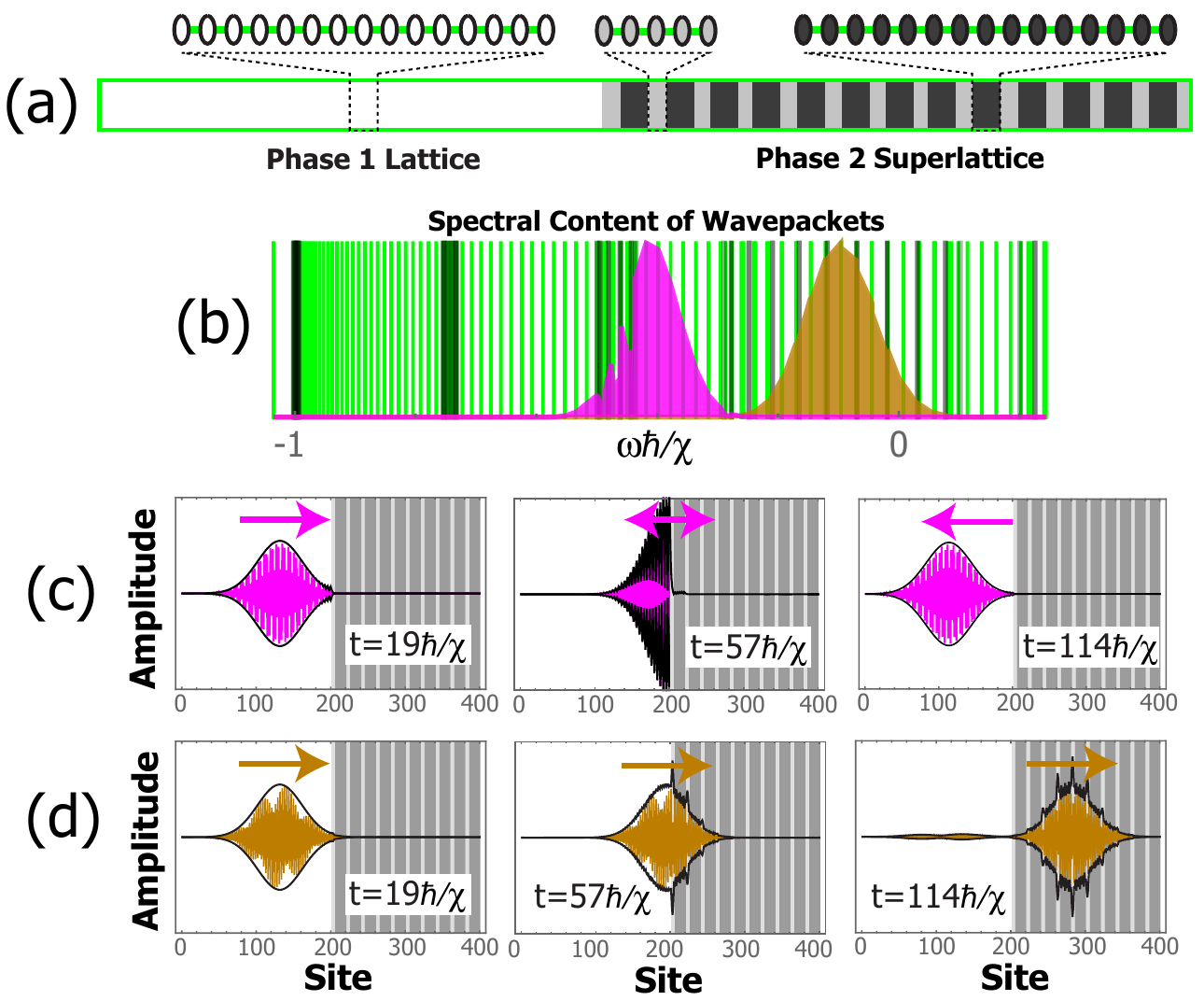}
\caption{
{\bf Excitonic Stop and Pass Bands.} {\bf (a)} A region of homogeneous sites (white, left) is joined to a superlattice composed of alternating crystalline layers (light/dark gray, right). {\bf (b)} The eigenvalues of the stand-alone superlattice (black lines) are plotted along with those of the entire system (green lines). An exciton is constructed that is composed of eigenvalues lying in a stop band (projection magnitudes in magenta) or a pass band (projection magnitudes in brown) of the superlattice. {\bf (c)} The stop band exciton (real part in magenta, absolute value in black) is plotted for three times to show that it is reflected at the phase boundary. {\bf (d)} The pass band exciton (real part in brown, absolute value in black) is plotted for three times to show that it is largely transmitted at the phase boundary. Here $N=400$,  $\Delta_1=\chi$, $\Delta_{2a}=0.5\chi$, $\Delta_{2b}=1.5\chi$ and $\sigma = 30.0 a$. There are 5 and 15 sites in the superlattice, and the central wavenumbers of the exciton are $k_0 = 2.36/a$ (panel c) $k_0 = 2.15/a$ (panel d).}
\label{Both_Bands}
\end{center}
\end{figure}

For instance, consider the  two-phase material (top schematics in Fig. \ref{Both_Bands}) in which the left phase is composed of a set of homogeneous sites and the right phase is a superlattice  of alternating layers with distinct energy gaps. An exciton traveling to the right will exhibit a transmission coefficient that can be anticipated from its central wavenumber. For example, if the exciton is constructed from eigenstates of the entire system with eigenvalues that lie in a stop band of the superlattice, then it will be effectively reflected, as shown in Fig.~\ref{Both_Bands}(c). This is because the wave function in the thin (lighter) layers is evanescent. On the other hand, an exciton in the same material and with same footprint, but constructed from eigenstates associated with a pass band of the superlattice, will be largely transmitted. This is shown in Fig.~\ref{Both_Bands}(d) and is the charge-neutral, excitonic analog of electron bandpass filtering. A wealth of information from ballistic electron experiments can be used to optimize their performance~\cite{Tung_1996}.

%
%

\begin{figure}[t]\begin{center}
\includegraphics[width=0.45\textwidth]{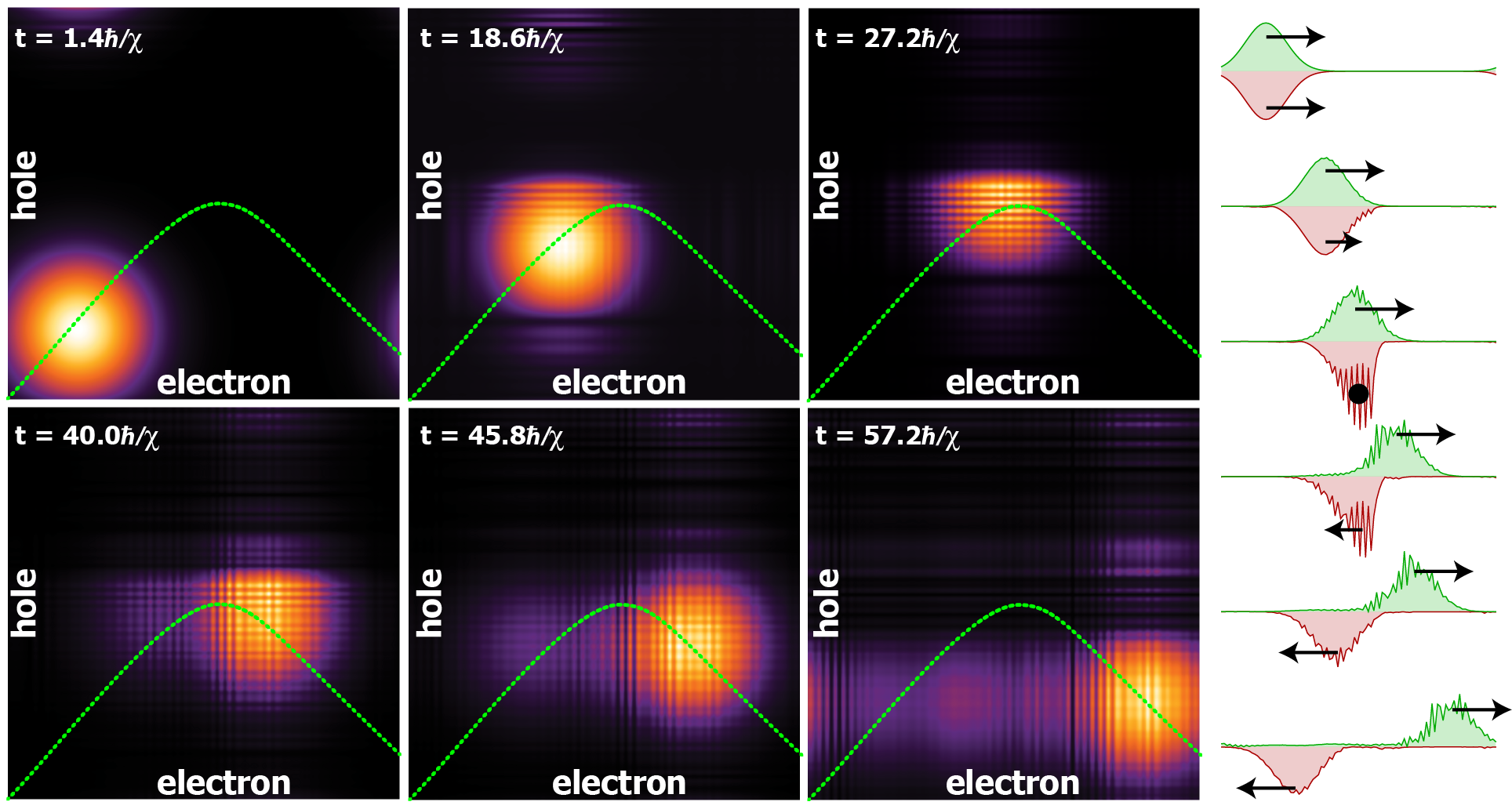} 
\caption{
{\bf Exciton Dissociation using Quantum Interference.}  Left panels show snapshots of a density plot of the probability distribution of the exciton along the chain. The green trajectory line is a guide to the eye. Right panels depict the same dynamics representing the electron (hole) probability projections in green (red) and the corresponding wave packet velocity for times shown at left (top to bottom). Here $N = 100$,  $\Delta_1^0 = 2\chi$, $\Delta_1^1 = 2\chi$, $\Delta_{2a}^0 = 0.0$, $\Delta_{2b}^0=2\chi$, $\Delta_{2a}^1 = 1.6\chi$, $\Delta_{2b}^1 = 2.4\chi$, $\chi$ is the same for both bands, $\sigma = 10.4 a$ and the Coulomb interactions have been turned off---the simplest setting for dissociation. There are 5 sites in each layer of phase 2, and the central wavenumber of the ring exciton is $k_0 = 1.63/a$.}
\label{Dissociation}
\end{center}
\end{figure}

The TB formalism can also be generalized to consider dynamics with distinct electron and hole dynamics using the Hamiltonian of Eq. \ref{H2}. This setting has been previously shown to allow Fano antiresonance to gate and dissociate excitons~\cite{Lusk_Fano_2015}, but they can also be dissociated into coherently linked, nonlocal electron and hole packets using a stop band filter. This is shown in Fig. \ref{Dissociation}, where the geometry of Fig. \ref{Both_Bands}(a) is adopted with the two-particle Hamiltonian of Eq. \ref{H2}.  

To demonstrate the idea, an exciton is constructed from states for which excited electrons lie in a pass band of the superlattice while ground state electrons (and so holes) lie in a stop band. As a result, an exciton that encounters the superlattice is largely dissociated into a spatially separated electron and a hole, as shown in Fig. \ref{Dissociation}. The right series of vertically arranged red and green time slices shows the evolution of the electron and hole probability densities for a Wannier-Mott exciton created by a laser pulse at the left end of the chain. The two-dimensional grids at left are for the same time slices but give additional detail by showing how the exciton is spatially distributed. Each grid point, $(i,j)$, in the NxN array gives the intensity of the exciton component for which the electron is at site $i$ and the hole is at site $j$. The exciton is initially created with an isotropic, two-dimensional distribution of states in the electron-hole population space. The electron and hole occupations are transmitted at nearly the same speed, highlighted with a green light at a positive 45-degree angle. The packet subsequently begins to interact with the superlattice when half-way through the chain of sites. This causes the hole to be reflected, seen as a downward motion of the packet in electron-hole space. However, the electron is transmitted so the packet continues to have a left-to-right motion. The result is a motion of the packet along the green-highlighted, negative 45-degree line.

%

\section{Exciton Stability}
This initial consideration of exciton wave packets has necessarily focused on their creation and manipulation, and the important influences of phonon entanglement and static/dynamic disorder have been neglected. In the absence of any phonons, though, even mild disorder can cause an exciton to localize~\cite{Anderson_1958} provided the system is larger than the relevant localization length. Likewise, interaction with phonons will result in a loss of coherence~\cite{Breuer_2002}. However, a degree of phonon entanglement breaks Anderson localization allowing partially coherent excitonic wave packets to propagate through regions of disorder~\cite{Plenio_2008, Kreisbeck_JCTC_2011}. Over the past several years, efforts to understand and exploit this behavior have focused on organic materials, and photosynthetic complexes in particular, in an effort to identify naturally occurring, long-lived quantum transport~\cite{Engel,  Collini_Science_2009}. Such biological settings, in which coherence is preserved for tens to hundreds of femtoseconds at room temperature, serve to inspire and guide the design of engineered materials and much colder temperature regimes for which entanglement can be tailored to make propagation robust in the face of disorder\cite{Lusk_PRB_2015}. 

Even for idealized systems in which disorder and entanglement are not issues, excitons will tend to broaden as they move. After all, it is just this dispersive nature that allows their speed to be tailored. The bandwidth of phase velocities is four times the hopping mobility, $\chi$, so decreasing the mobility would reduce dispersion but at the cost of less tunable exciton speeds and excitons whose slower speed leaves them more susceptible to decoherence. A more practical approach is to increase the spatial footprint of the exciton. This narrows the energy range, and so speed disparities, of the modes of which they are comprised. This philosophy was used in producing the results shown in Fig. ~\ref{Packet_Evolution}, where dispersion is seen to be quite low. 

\section{Conclusions}

Structured exciton wave packets offer the prospect of fabricating opto-excitonic circuits in which photon energy/information is easily processed and transmitted as excitons. These excitons can be selectively gated, subjected to filters and even dissociated. A methodology for creating and manipulating such excitons has been demonstrated using idealized Tight-Binding models and also more realistic Time-Domain Density Functional Theory simulations. 

The associated exciton circuit embodies many of the key properties of Heisenberg spin chains~\cite{Bose_PRL_2003, Bose_ContempPhys_2007,Thompson_2016} that are often considered in association with quantum information processing~\cite{Wang_PRA_2012}. The focus there tends to be on the high-quality transmission of data down \emph{quantum buses}, and several non-excitonic experimental implementations of spin chains now exist~\cite{You_Nature_2011,Qin_PRA_2013, Ping_PRL_2013, Farooq_PRB_2015} with particles such as phonons, electrons, photons, magnons, and Cooper pairs. All of ideas associated with such spin chains can be incorporated into the current paradigm~\cite{Lukin_PRA_2002, Lukin_PRA_2013, Caneva_PRL_2009, Marvian_PRL_2015}. It is also intriguing to consider what impact exciton wave packets might have on the design of light-harvesting complexes and on optimally balancing their entanglement with phonons.

Perhaps the most controllable setting for initial exploration of controllable, ballistic excitons is a chain of Rydberg atoms trapped in an optical lattice~\cite{Glaetzle2015a, Zeiher_2015}. Chains of dressed Rydberg atoms~\cite{Wuster_NJP_2011} can now be produced that support exciton-like states, and computational models indicate that it is possible to use pulsed magnetic fields to create rudimentary, traveling superpositions with a degree of localization~\cite{Schempp_PRL_2015}. Strongly coupled optical cavities with encapsulated excitonic structures offer a comparable setting with the additional benefit of exquisite control over light/matter coupling~\cite{Yariv_1999, Ergecen_2014}.

It was assumed that Two-Photon Absorption can be used to excite isolated end sites without adversely affecting excitonic coupling to the rest of the system. The prescribed electric field is treated classically, and two-way coupling between light and matter~\cite{MSBE_2011} is disregarded because the materials are optically thin \cite{Haug_2009}. The introduction of exciton-phonon coupling would result in partially coherent transport which is more robust in the face of material imperfections~\cite{Plenio_2008, Arago_AFM_2015} but must be managed in order to preserve the basic phase relationships between eigenstates. Idealized molecular chains, consisting of benzene and methane sites, were use to demonstrate the possibility of creating and transmitting exciton wave packets. This is not intended to represent a recommended material choice for experimental implementation because of the high excitation energies, low mobilities, and no obvious means of creating such a chain.  

Exciton-exciton interactions have also been neglected in this investigation, but they offer the prospect of incorporating the many nonlinear effects associated with photonic crystals within a new physical setting. Particularly in this regard, excitonic circuits are analogous to photonic crystals and coupled optical cavities, where particle-particle interactions are a central focus. The structure of excitons can also be broadened to include spin engineering. This has been disregarded here, but it may prove interesting in creating delocalized, entangled states when lattice interactions result in both transmitted and reflected exciton components or spatially separated electron and hole superpositions.

Excitonic superlattices, and the associated stop/pass bands, can be used to create entangled states between reflected and transmitted exciton packets or between electron and hole packets that have been dissociated. Laser pulse engineering thus opens the door to the prospect of studying EPR-like phenomena within an excitonic setting. 

Finally, more complex protocols could be engineered considering multi-site excitation, exciton-exciton interactions, phonon coupling, finite temperature and higher number of excitons, by means of quantum optimal control methods also combined with RT-TDDFT or Tensor Network methods~\cite{Doria2011}.

\begin{acknowledgments}
SM acknowledges support from the DFG (German Research Foundation) via the SFB/TRR21 and the EU via the SIQS and RYSQ projects. This material is based in part upon work supported by the National Science Foundation under grant numbers PHY-1306638, PHY-1207881, PHY-1520915, and the Air Force Office of Scientific Research grant number FA9550-14-1-0287. All computations were carried out using the High Performance Computing facilities at the Colorado School of Mines. S.M. gratefully acknowledges the support of the DFG via a Heisenberg fellowship.
\end{acknowledgments}


\begin{thebibliography}{10}

\bibitem{Mikhnenko_2015}
O.~V. Mikhnenko, P.~W.~M. Blom, and T.-Q. Nguyen, Energy Environ. Sci. {\bf 8},
   1867  (2015).

\bibitem{Hahn_PRB_1980}
E.~L. Hahn, Phys. Rev. {\bf 80},  580  (1950).

\bibitem{Stolz_PRL_1991}
H. Stolz, V. Langer, E. Schreiber, S. Permogorov, and W. von~der Osten, Phys.
  Rev. Lett. {\bf 67},  679  (1991).

\bibitem{Feldmann_1993}
J. Feldmann, T. Meier, G. von Plessen, M. Koch, E.~O. G\"obel, P. Thomas, G.
  Bacher, C. Hartmann, H. Schweizer, W. Sch\"afer, and H. Nickel, Phys. Rev.
  Lett. {\bf 70},  3027  (1993).

\bibitem{Elsaesser_PRL_1991}
T. Elsaesser, J. Shah, L. Rota, and P. Lugli, Phys. Rev. Lett. {\bf 66},  1757
  (1991).

\bibitem{Bayer_Science_2001}
M. Bayer, P. Hawrylak, K. Hinzer, S. Fafard, M. Korkusinski, Z. Wasilewski, O.
  Stern, and A. Forchel, Science (New York, N.Y.) {\bf 291},  451  (2001).

\bibitem{Collini_Science_2009}
E. Collini and G.~D. Scholes, Science {\bf 323},  369  (2009).

\bibitem{Delbecq_NatComm_2013}
M.~R. Delbecq, L.~E. Bruhat, J.~J. Viennot, S. Datta, A. Cottet, and T. Kontos,
  Nat Commun {\bf 4},  1400  (2013).

\bibitem{Rivas}
D. Rivas, G. Muñoz-Matutano, J. Canet-Ferrer, R. García-Calzada, G. Trevisi,
  L. Seravalli, P. Frigeri, and J.~P. Martínez-Pastor, Nano Lett. {\bf 14},
  456  (2014).

\bibitem{Leisching_PRB_1994}
P. Leisching, P. Haring~Bolivar, W. Beck, Y. Dhaibi, F. Br\"uggemann, R.
  Schwedler, H. Kurz, K. Leo, and K. K\"ohler, Phys. Rev. B {\bf 50},  14389
  (1994).

\bibitem{Yang_APL_2014}
S. Yang, X. Tian, L. Wang, J. Wei, K. Qi, X. Li, Z. Xu, W. Wang, J. Zhao, X.
  Bai, and E. Wang, Applied Physics Letters {\bf 105},  071901  (2014).

\bibitem{Jang}
S. Jang, Y.-C. Cheng, D.~R. Reichman, and J.~D. Eaves, J. Chem. Phys. {\bf
  129},  101104  (2008).

\bibitem{Kreisbeck_2012}
C. Kreisbeck and T. Kramer, The Journal of Physical Chemistry Letters {\bf 3},
  2828  (2012).

\bibitem{Plenio_2008}
M.~B. Plenio and S.~F. Huelga, New Journal of Physics {\bf 10},  113019
  (2008).

\bibitem{Umlauff_1998}
M. Umlauff, J. Hoffmann, H. Kalt, W. Langbein, J.~M. Hvam, M. Scholl, J.
  S\"ollner, M. Heuken, B. Jobst, and D. Hommel, Phys. Rev. B {\bf 57},  1390
  (1998).

\bibitem{Kalt_2005}
H. Kalt, H. Zhao, B.~D. Don, G. Schwartz, C. Bradford, and K. Prior, Journal of
  Luminescence {\bf 112},  136   (2005).

\bibitem{Yablonovitch_PRL_1987}
E. Yablonovitch, Phys. Rev. Lett. {\bf 58},  2059  (1987).

\bibitem{Krauss_Nature_1996}
T.~F. Krauss, R.~M. D.~L. Rue, and S. Brand, Nature {\bf 383},  699  (1996).

\bibitem{Gartner}
A. G${\rm \ddot a}$rtner, A.~W. Holleitner, J.~P. Kotthaus, and D. Schuh, Appl.
  Phys. Lett. {\bf 89},  052108  (2006).

\bibitem{Wuster_NJP_2011}
S. Wuster, C. Ates, A. Eisfeld, and J.~M. Rost, New Journal of Physics {\bf
  13},  073044  (2011).

\bibitem{Lusk_Fano_2015}
M.~T. Lusk, C.~A. Stafford, J.~D. Zimmerman, and L.~D. Carr, Phys. Rev. B {\bf
  92},  241112  (2015).

\bibitem{Osborne_PRA_2004}
T.~J. Osborne and N. Linden, Phys. Rev. A {\bf 69},  052315  (2004).

\bibitem{Haselgrove_PRA_2005}
H.~L. Haselgrove, Phys. Rev. A {\bf 72},  062326  (2005).

\bibitem{Bose_ContempPhys_2007}
S. Bose, Contemporary Physics {\bf 48},  13  (2007).

\bibitem{Thompson_2016}
K.~F. Thompson, C. Gokler, S. Lloyd, and P.~W. Shor, New Journal of Physics
  {\bf 18},  073044  (2016).

\bibitem{Seifnashri_2016}
S. Seifnashri, F. Kianvash, J. Nobakht, and V. Karimipour, Phys. Rev. A {\bf
  93},  062342  (2016).

\bibitem{Ishizaki_2009}
A. Ishizaki and G.~R. Fleming, The Journal of Chemical Physics {\bf 130},
  (2009).

\bibitem{Kaiser_1961}
W. Kaiser and C.~G.~B. Garrett, Phys. Rev. Lett. {\bf 7},  229  (1961).

\bibitem{Rumi_2010}
M. Rumi and J.~W. Perry, Adv. Opt. Photon. {\bf 2},  451  (2010).

\bibitem{Marder_2006}
S.-J. Chung,  . Shijun~Zheng, T. Odani, L. Beverina, €. Jie~Fu, L.~A. Padilha,
  A. Biesso, J.~M. Hales, X. Zhan, K. Schmidt, A. Ye, E. Zojer, S. Barlow,
  D.~J. Hagan, E.~W.~V. Stryland, Y. Yi, Z. Shuai, G.~A. Pagani, J.-L. Bredas,
  J.~W. Perry , and S.~R. Marder, Journal of the American Chemical Society {\bf
  128},  14444  (2006), pMID: 17090012.

\bibitem{Hu_2013}
L. Hu, Z. Yan, and H. Xu, RSC Adv. {\bf 3},  7667  (2013).

\bibitem{Schempp_PRL_2015}
H. Schempp, G. G\"unter, S. W\"uster, M. Weidem\"uller, and S. Whitlock, Phys.
  Rev. Lett. {\bf 115},  093002  (2015).

\bibitem{Runge_1984}
E. Runge and E.~K.~U. Gross, Phys. Rev. Lett. {\bf 52},  997  (1984).

\bibitem{Peng_2015}
B. Peng, D.~B. Lingerfelt, F. Ding, C.~M. Aikens, and X. Li, The Journal of
  Physical Chemistry C {\bf 119},  6421  (2015).

\bibitem{Xavier_2015}
X. Andrade, D. Strubbe, U. De~Giovannini, A.~H. Larsen, M.~J.~T. Oliveira, J.
  Alberdi-Rodriguez, A. Varas, I. Theophilou, N. Helbig, M.~J. Verstraete, L.
  Stella, F. Nogueira, A. Aspuru-Guzik, A. Castro, M.~A.~L. Marques, and A.
  Rubio, Phys. Chem. Chem. Phys. {\bf 17},  31371  (2015).

\bibitem{Cong2012}
C. Wang, L. Jiang, F. Wang, X. Li, Y. Yuan, H. Xiao, H.-L. Tsai, and Y. Lu,
  Journal of Physics: Condensed Matter {\bf 24},  275801  (2012).

\bibitem{Yabana2012}
K. Yabana, T. Sugiyama, Y. Shinohara, T. Otobe, and G.~F. Bertsch, Phys. Rev. B
  {\bf 85},  045134  (2012).

\bibitem{Lopata2011}
K. Lopata and N. Govind, Journal of Chemical Theory and Computation {\bf 7},
  1344  (2011).

\bibitem{Hellmann_1937}
Clusius, Angewandte Chemie {\bf 54},  156  (1941).

\bibitem{Feynman_1939}
R.~P. Feynman, Phys. Rev. {\bf 56},  340  (1939).

\bibitem{AD}
M. Head-Gordon, A.~M. Grana, D. Maurice, and C.~A. White, J Phys. Chem. {\bf
  99},  14261  (1995).

\bibitem{Octopus_2006}
A. Castro, H. Appel, M. Oliveira, C.~A. Rozzi, X. Andrade, F. Lorenzen,
  M.~A.~L. Marques, E.~K.~U. Gross, and A. Rubio, physica status solidi (b)
  {\bf 243},  2465  (2006).

\bibitem{FED1}
C.-P. Hsu, Z.-Q. You, and H.-C. Chen, The Journal of Physical Chemistry C {\bf
  112},  1204  (2008).

\bibitem{FED2}
H.-C. Chen, Z.-Q. You, and C.-P. Hsu, The Journal of Chemical Physics {\bf
  129},  084708  (2008).

\bibitem{FED3}
C.-P. Hsu, Accounts of Chemical Research {\bf 42},  509  (2009).

\bibitem{Subotnik2}
J.~E. Subotnik, R.~J. Cave, R.~P. Steele, and N. Shenvi, The Journal of
  Chemical Physics {\bf 130},  234102  (2009).

\bibitem{ER}
C. Edmiston and K. Ruedenberg, Rev. Mod. Phys. {\bf 35},  457  (1963).

\bibitem{Tung_1996}
H.-H. Tung and C.-P. Lee, IEEE Journal of Quantum Electronics {\bf 32},  507
  (1996).

\bibitem{Anderson_1958}
P.~W. Anderson, Phys. Rev. {\bf 109},  1492  (1958).

\bibitem{Breuer_2002}
H.-P. Breuer and F. Petruccione, {\em The theory of open quantum systems}
  (Oxford University Press, Oxford New York, 2002).

\bibitem{Kreisbeck_JCTC_2011}
C. Kreisbeck, T. Kramer, M. Rodríguez, and B. Hein, Journal of Chemical Theory
  and Computation {\bf 7},  2166  (2011).

\bibitem{Engel}
G.~S. Engel, T.~R. Calhoun, E.~L. Read, T.-K. Ahn, T. Manc$\breve{a}$, Y.~C.
  Cheng, R.~E. Blankenship, and G.~R. Fleming, Nature {\bf 446},  782   (2007).

\bibitem{Lusk_PRB_2015}
X. Zang and M.~T. Lusk, Phys. Rev. B {\bf 92},  035426  (2015).

\bibitem{Bose_PRL_2003}
S. Bose, Phys. Rev. Lett. {\bf 91},  207901  (2003).

\bibitem{Wang_PRA_2012}
Z.-M. Wang, R.-S. Ma, C.~A. Bishop, and Y.-J. Gu, Phys. Rev. A {\bf 86},
  022330  (2012).

\bibitem{You_Nature_2011}
J.~Q. You and F. Nori, Nature {\bf 474},  589  (2011), 10.1038/nature10122.

\bibitem{Qin_PRA_2013}
W. Qin, C. Wang, and G.~L. Long, Phys. Rev. A {\bf 87},  012339  (2013).

\bibitem{Ping_PRL_2013}
Y. Ping, B.~W. Lovett, S.~C. Benjamin, and E.~M. Gauger, Phys. Rev. Lett. {\bf
  110},  100503  (2013).

\bibitem{Farooq_PRB_2015}
U. Farooq, A. Bayat, S. Mancini, and S. Bose, Phys. Rev. B {\bf 91},  134303
  (2015).

\bibitem{Lukin_PRA_2002}
M. Fleischhauer and M.~D. Lukin, Phys. Rev. A {\bf 65},  022314  (2002).

\bibitem{Lukin_PRA_2013}
N.~Y. Yao, Z.-X. Gong, C.~R. Laumann, S.~D. Bennett, L.-M. Duan, M.~D. Lukin,
  L. Jiang, and A.~V. Gorshkov, Phys. Rev. A {\bf 87},  022306  (2013).

\bibitem{Caneva_PRL_2009}
T. Caneva, M. Murphy, T. Calarco, R. Fazio, S. Montangero, V. Giovannetti, and
  G.~E. Santoro, Phys. Rev. Lett. {\bf 103},  240501  (2009).

\bibitem{Marvian_PRL_2015}
I. Marvian and D.~A. Lidar, Phys. Rev. Lett. {\bf 115},  210402  (2015).

\bibitem{Glaetzle2015a}
A.~W. Glaetzle, M. Dalmonte, R. Nath, C. Gross, I. Bloch, and P. Zoller, Phys.
  Rev. Lett. {\bf 114},  173002  (2015).

\bibitem{Zeiher_2015}
J. Zeiher, P. Schau\ss{}, S. Hild, T. Macr\`{\i}, I. Bloch, and C. Gross, Phys.
  Rev. X {\bf 5},  031015  (2015).

\bibitem{Yariv_1999}
A. Yariv, Y. Xu, R.~K. Lee, and A. Scherer, Opt. Lett. {\bf 24},  711  (1999).

\bibitem{Ergecen_2014}
{Ergecen, Emre},  in {\em {Nonlinear Optics and Its Applications VIII; And
  Quantum Optics III}}, Vol.~{9136} of {\em {Proceedings of SPIE}}, edited by
  {Eggleton, BJ and Gaeta, AL and Broderick, NGR and Sergienko, AV and
  Rauschenbeutel, A and Durt, T} ({SPIE-International Society of Optical
  Engineering}, ADDRESS, {2014}).

\bibitem{MSBE_2011}
{\em Semiconductor Quantum Optics:} (Cambridge University Press, ADDRESS,
  2011), pp.\ 521--549.

\bibitem{Haug_2009}
H. Haug, {\em Quantum theory of the optical and electronic properties of
  semiconductors} (World Scientific, Singapore Hackensack, N.J, 2009).

\bibitem{Arago_AFM_2015}
J. Arag$\acute{o}$ and A. Troisi, Advanced Functional Materials {\bf 26},  1
  (2015).

\bibitem{Doria2011}
P. Doria, T. Calarco, and S. Montangero, Phys. Rev. Lett. {\bf 106},  190501
  (2011).

\end{thebibliography}

\end{document}